\documentclass[fleqn,usenatbib]{mnras}

\usepackage{newtxtext,newtxmath}
\usepackage{ulem}
\usepackage{cancel}
\usepackage[T1]{fontenc}

\DeclareRobustCommand{\VAN}[3]{#2}
\let\VANthebibliography\thebibliography
\def\thebibliography{\DeclareRobustCommand{\VAN}[3]{##3}\VANthebibliography}

\usepackage{graphicx}	
\usepackage{amsmath}	
\usepackage{xcolor}

\newcommand{\Bete}[1]{\textcolor{red}{#1}}

\title[
Multimessenger emission from reconnection in blazars]{
Multi-messenger emission from magnetic reconnection in blazar jets: the case of TXS 0506+056}

\author
[E. M. de Gouveia Dal Pino et al.]{
E. M. de Gouveia Dal Pino$^{1}$,\thanks{E-mail: dalpino@iag.usp.br, juancr@cbpf.br}
J. C. Rodr\'iguez-Ram\'irez$^{1,2}$,
M. V. del Valle$^{1}$
\\
$^{1}$
Instituto de Astronomia, Geof\'isica e 
Ci\^{e}ncias Atmosf\'ericas
Universidade de S\~ao Paulo IAG,
Rua do Mat\~ao 1225,
CEP: 05508-090
S\~ao Paulo - SP - Brazil.\\
$^{2}$Centro Brasileiro de Pesquisas F\'isicas,
Rua Dr. Xavier Sigaud, 150, CEP: 22290-180, 
Rio de Janeiro - RJ - Brazil.
}

\date{Accepted XXX. Received YYY; in original form ZZZ}

\pubyear{2023}

\begin{document}
\label{firstpage}
\pagerange{\pageref{firstpage}--\pageref{lastpage}}
\maketitle


\begin{abstract}

Measurements from astroparticle experiments, such as the 2017 flare associated with the source TXS 0506+056, indicate that blazars act as multi-messenger (MM; radiation and neutrinos) factories. Theoretically, the particle acceleration mechanisms responsible for blazar emissions and the precise location within the jet where this occurs remain undetermined.
This paper explores MM emission driven by magnetic reconnection in a blazar jet. 
Previous studies have shown that reconnection in the magnetically dominated regions of these relativistic jets can efficiently accelerate particles to very high energies (VHE). Assuming that turbulent-driven magnetic reconnection accelerates cosmic-ray protons and electrons by a Fermi process, we developed a lepto-hadronic radiation model without the influence of external soft-photons  to explain the 2017 MM flare from TXS 0506+056. In the proposed scenario, the emission blob moves downstream in the jet from $\sim$2 to  4 pc  from the central engine, which is a supermassive black hole (SMBH) of $3 \times 10^{8}$ M$_\odot$ launching a jet with $150L_\mathrm{Edd}$ power. As the blob moves, we observe a sequence of spectral energy distribution (SED) profiles that match the observed arrival of the high energy neutrino and electromagnetic emission from TXS 0506+056. This arrival coincides with the high state of intermediate energy $\gamma$-rays ($E \sim 1 $ GeV) detection, followed by the subsequent appearance of the VHE $\gamma$-ray signal and then  no further significant neutrino detection. We obtain a time delay between the neutrino and VHE events  $\simeq 6.4$ days, which is  consistent with that observed  in the 2017 MM flare.
\\

\end{abstract}




\section{Introduction}
\label{sec:intro}

Relativistic jets are collimated outflows of plasma propagating at relativistic speeds, commonly observed in active galaxies and, on a much smaller scale, in our own galaxy in sources known as microquasars. These high-energy outflows are generated and energized by supermassive black holes (SMBHs) and stellar-mass black holes, respectively. Even more extreme cases occur in gamma-ray bursts, which are powered by the collapse of compact stars into black holes \citep[e.g.][and references therein]{dalpino_etal_2020}.
One of the most accepted explanations for the origin of such energetic collimated outflows  is the Blandford-Znajek mechanism \citep{1977MNRAS.179..433B,2012MNRAS.423.3083M}. In this scenario, relativistic jets 
are launched by magnetic torques of accumulated field lines anchored to the highly spinning BHs.
If this description is accurate, relativistic jets are  expected  to be born magnetically dominated and hence they must convert magnetic to  kinetic energy to achieve their observed relativistic speeds and emissions \citep[e.g.][for reviews]{giannios_etal_09, giannios_10, dalpino_etal_2020}


Functioning as highly efficient particle accelerators, these jets frequently emit radiation at GeV and TeV energies \citep[see e.g. review by][]{hovatta_lindfors19}.
Recent observations have provided compelling evidence of these jets' ability to accelerate protons, as demonstrated by the case of TXS 0506+056, where simultaneous emissions of high-energy gamma-rays and neutrino were detected \citep{neutrinosmm}. 
However, ongoing discussions within the scientific community raise questions regarding the extent to which these jets contribute to the observed astrophysical neutrinos (both diffuse and localized) recorded by instruments like IceCube \citep[e.g.][]{Khiali2016, neutrinosprior,neutrinosmm,murae2018, Plavin2020, Plavin2021,  Hovatta2021, Abbasi2022, Buson2022, Bellenghi2023}.

Recently, in light of ongoing debates on the origin of cosmic ray (CR) acceleration and very high energy (VHE) variable emission, particularly in blazars 
\citep[e.g.][]{aharonian_etal_07, ackermann_etal_2016, britto_elal_2016,  neutrinosprior, neutrinosmm},
extensive studies have been conducted on particle acceleration by magnetic reconnection in magnetically dominated regions where shocks are faint or absent. 
Our understanding of particle acceleration driven by magnetic reconnection has significantly advanced due to both particle-in-cell (PIC) simulations predominantly performed in two-dimensions (2D) \citep[e.g.][]{lyubarsky_etal_2008,    li_etal_2015, petropoulou_etal_2016, werner_etal_2019,
lyutikov_etal_2017, sironi_spitkovsky_2014, guo_etal_2020,  kilian_etal_2020, comisso18, cerruti_2020,  guo_etal_2022}, 
and 3D magnetohydrodynamic (MHD) simulations 
\citep[e.g.][]{kowal_etal_2011, kowal_etal_2012, dalpino_kowal_15, delvalle_etal_16, beresnyak_etal_2016, Ripperda2017, kadowaki_etal_2021, Medina-Torrejón_2021, Zhang_Xu2023, 
Medina-Torrejón_2023}. 
While PIC simulations probe the micro-scales of the process, studying acceleration up to at most $10^4$ times the particle's rest mass, MHD simulations with test particles probe the macro-scales, detecting particle acceleration up to observed VHE and ultra-high energies (UHECRs) \citep[][]{Medina-Torrejón_2021, Medina-Torrejón_2023}. 


In particular, \cite{Medina-Torrejón_2021}, 
\cite{kadowaki_etal_2021}, 
and \cite{Medina-Torrejón_2023}, 
conducted 3D MHD and 3D MHD-PIC simulations exploring how relativistic jets transition from magnetically to kinetically dominated states with a magnetization parameter (given by the ratio between the magnetic and the rest mass energy density) close to 1. They observed that jets, affected by current-driven kink instability become turbulent, forming  multiple fast reconnection layers at all scales of the turbulent flow \citep[as predicted by][theory]{lazarian_vishiniac_99}. Test protons injected into this turbulent flow with reconnection layers undergo  acceleration with nearly exponential energy growth in time, primarily in their parallel momentum component, akin to a Fermi process \citep[][]{dalpino_lazarian_2005}.
 They reach threshold energies, with Larmor radius as large as the size of the jet diameter, before undergoing further, slower acceleration drifting in non-reconnecting magnetic fields. 
Other studies involving PIC simulations of relativistic jets subject to kink instability  \citep{davelaar_etal_2020}
and  other driving mechanisms of turbulence such as Weibel, kinetic Kelvin-Helmholtz, and mushroom instability \citep{nishikawa_etal_2020}, have also detected particle acceleration by magnetic reconnection, though to much smaller energies, as allowed by the small scales probed by kinetic (PIC) simulations. These findings underscore the potential of magnetic reconnection to drive particle acceleration, potentially explaining UHE phenomena in blazars \citep[see also][]{giannios_10, dalpino_etal_2020}.
Whether or not this process  produces observable multi-messenger emissions can be tested, for the first time,   using the 2017 multi-messenger flare from  TXS 0506+056.

TXS 0506+056 is a well-known blazar 
localized at redshift $z = 0.34$ \citep{2018ApJ...854L..32P}.  A high-energy neutrino event, a moun neutrino with $E\sim 290$\,TeV (IceCube-170922A), detected by IceCube on 22 September 2017 was coincident in direction and time with a gamma-ray flare from this source \citep{neutrinosmm}.  Analysis of 9.5 years archival IceCube data revealed that this blazar was emitting neutrinos between October 2014 and March 2015. An  excess of 13$\pm$5 events above the expectation from the atmospheric background was found. In opposition to IceCube-170922A, no electromagnetic flaring activity was present during these neutrino events \citep{neutrinosprior}. Both occurrences prove blazars as the first identifiable sources of the high-energy astrophysical neutrino flux.


Explaining the 13$\pm$5 neutrinos observed during the 2014–2015 period has proven challenging. Previous studies \citep[][]{reimer2019, rodrigues2019} have shown that even under optimistic assumptions regarding the emitting region and the photon target field (whether internal or external to the jet), only a few neutrino events can be reproduced. Achieving consistency between the observed multiwavelength (MW) data and the neutrino detections would require two-zone models, introducing additional free parameters.
Several scenarios have been proposed to address both the 2014–2015 neutrino excess and the IC-170922A neutrino episodes, such as neutral beams \citep[][]{zhang_etal_2018}, precessing jet-jet interactions \citep[][]{britzen_2019} or more general time-dependent models \citep[][]{gasparian_2022}. These have aimed to link both neutrino episodes to TXS 0506+056. However, the complexity and uncertainties associated with explaining both neutrino episodes have led many studies to focus specifically on the IC-170922A neutrino event and its association with the 2017 MW flaring state of TXS 0506+056. This targeted approach has allowed for more detailed exploration of the acceleration and emission mechanisms during the 2017 flare.

Under the hypothesis that the IC-170922A neutrino and the 2017 MW flaring state of TXS 0506+056 both originated in the blazar jet, multiple  radiative approaches have been explored to explain the multi-messenger (MM) data. 
Considering the photo-pion reaction as the principal channel for neutrino production in the jet, independent studies appear to agree that mixed (also called hybrid) lepto-hadronic scenarios lead to higher rates of neutrinos production compared to proton synchrotron models at energies consistent with the IC-170922A event \citep{Cerruti_2019,Keivani_2018,Gao_2019,petropoulou2020, gasparian_2022}.
Regarding the source of target photons for photo-pion production,
the simplest scenarios assume that the target photons are synchrotron photons from primary electrons within the jet, which are the same photons that produce the low-energy bump in the observed SED \citep{Cerruti_2019,Gao_2019}.
Models assisted with external sources of radiation, such as a putative broad line region (BLR)  
\citep{Keivani_2018,Xue_2019}, 
or an external sheath region of a structured jet \citep{Ansoldi_2018,zhang_murase2020}, can also in principle explain the  MM  SED.
Although scenarios with external target photons add complexity to the radiation model (as it is not clear yet where the photons come from 
in TXS 0506+056), they have the characteristic of demanding less jet power compared to models with only internal soft photons.
High-energy neutrinos can also be produced through the proton-proton ($pp$) hadronic channel \citep[e.g.][]{banik2019}. However, the rate of $pp$ interactions in blazar jets is generally negligible compared to the rate of photo-pion production, unless external mechanisms are present.
Under this line, the MM flare from TXS 0506+056 has also been investigated as a blazar jet interacting with an external obstacle (such as a BLR cloud or a star envelope \citep{Sahakyan_2018,Liu_2019}. Still, the lack of broadline signatures also creates doubt about the presence of a high density obstacle in the vicinity of the black hole of TXS 0506 +056.

In this paper, we focus on an yet less explored aspect:  the acceleration mechanism associated with the MM emission. 
We propose that the MM 
emission from the 2017 flare of TXS 0506+056 is driven by magnetic reconnection in the blazar jet. To explore the maximum possible neutrino production at energy levels consistent with the IC-170922A event, we adopt a hybrid lepto-hadronic model, assuming the absence of external sources of soft radiation for the simplest possible radiative scenario.
In our model, we allow the large-scale properties of the blazar jet to transition from a magnetically dominated to a kinetically dominated flow as it propagates.
We then derive MM emission at the jet locations where magnetic reconnection is likely the operating mechanism of particle acceleration.
With this approach, as the emission region  moves downstream with the jet flow, it produces a sequence of SEDs that are able to reproduce the 2017 MM flare from the blazar TXS-0506+056. 

The paper is organised as follows.
In  section \ref{sec:striped_model}, we present the analytic model that we
adopt to parametrise the macroscopic properties of the jet as a function of the distance from the central black hole.
In Section~\ref{sec:rad_model}, we describe the basic assumptions for deriving the MM emission from the jet.
We apply this emission scenario, based on particle acceleration by magnetic reconnection within the emission region, to interpret the SED
of the 2017 MM flare from the source TXS-0506+056 in Section~\ref{sec:app_to_txs}.
Finally, in Section~\ref{sec:summary} we summarise
the present study and list our concluding  remarks.

\section{
Magnetic dissipation through the jet flow}

\label{sec:striped_model}

The numerical simulations involving particle acceleration by magnetic reconnection described above that were able to probe the production of very high energy cosmic rays have  a very complex environment with a large scale helical magnetic field mixed with a turbulent component \citep[e.g.][]{Medina-Torrejón_2021, Medina-Torrejón_2023}. In this background, particles are accelerated in the reconnection layers (or current sheets) driven by the turbulence. In order to build a similar but simpler scenario able to provide an analytical description of the development of fast reconnection and magnetic energy dissipation to energize  particles and allow for MM emission, we here adopt a simpler picture, which is described below. 


\cite{2019MNRAS.484.1378G} (hereafter GU19) have modelled the 
transition from magnetically  to kinetically dominated  regimes in the
flow of a relativistic jet due to the
magnetic reconnection  of  fields of opposite polarity normal to the jet axis, referred as stripes. Magnetic reconnection is regarded as the most  efficient  mechanism to allow magnetic energy dissipation into heating, kinetic energy as well as particle acceleration.
In their scenario, the magnetic stripes 
originate in the accretion disk where reversed fields are due to turbulent dynamo driven by the magneto-rotational instability (MRI). As they emerge, the stripes pile up along the jet (perpendicular to the axis), with a distribution of sizes, in a  similar, but much simpler configuration than that of the simulations in \cite{Medina-Torrejón_2021, Medina-Torrejón_2023}.

GU19 propose that the width $l$ of the stripes in the jet follows
a power-law distribution $\frac{dP}{dl} = \frac{1}{(a-1)l_\mathrm{min}}
(l/l_\mathrm{min})^{-a}$, being $l_\mathrm{min}$ the minimum stripe width, and 
that, e.g.,  tearing mode instability  induces magnetic reconnection 
between the stripes in the jet. 
As stressed in Section \ref{sec:intro}, in a  realistic scenario, the presence of turbulence driven by any instability that naturally develops in jet, such as kink, Kelvin-Helmholtz, Weibel, and even the MRI in the disk, may drive fast reconnection in the magnetized flow 
\citep[][]{kowal_etal_09, Kadowaki_2018,  nishikawa_etal_2020, kowal_etal_2020, Medina-Torrejón_2021,kadowaki_etal_2021, Medina-Torrejón_2023}. 
In recent work, very high resolution 3D MHD resistive simulations have demonstrated that   turbulence is  more efficient to drive fast reconnection  than tearing mode instability \citep[][]{Vicentin2024, Morillo2024}. Therefore, in this work, we  assume that fast reconnection between the stripes is driven by  embedded turbulence 
(see also section \ref{sec:acctime}).

Considering a conical jet with half opening angle
$\theta_\mathrm{j}$ and
assuming conservation of the total jet power 
$L_\mathrm{j} = L_\mathrm{B} + L_\mathrm{K}$,
where $L_\mathrm{B}$ and $L_\mathrm{K}$
are the magnetic and kinetic power, respectively,
GU19 derive an analytic model
describing 
the jet bulk motion that converges to an asymptotic Lorentz factor $\Gamma_\infty$.
We adopt this  formalism 
to 
asses the physical properties of a jet
dissipating magnetic energy by reconnection along the 
jet flow. 
Based on this jet model, we parametrise the macroscopic
properties  within the jet. 
We assume that particle acceleration by reconnection occurs at the jet's transition region, where it shifts from being magnetically dominated to kinetically dominated, as this is where the most significant energy emission and dissipation are expected.
The solutions of the GU19 model give
the magnetic dissipation power
$P_\mathrm{diss}$, and the jet bulk Lorentz factor $\Gamma_\mathrm{j}$ (both in the BH frame), at a 
distance $s$ from the BH,  as:
\begin{equation}
P_\mathrm{diss}(s) = 
L_\mathrm{j} \frac{\left[1-\chi(\zeta)\right]^{k}}{\chi^2(\zeta)}
\zeta,
\label{P_diss}
\end{equation}
\begin{equation}
\Gamma_\mathrm{j}(s) = \Gamma_\infty
\chi(\zeta),
\label{Gamma_j}
\end{equation}
where
\begin{equation}
\label{zeta}
\zeta = \frac{2\xi_\mathrm{rec} s}{l_\mathrm{min}\Gamma_\infty^{2}},
\end{equation}
and
\begin{equation}
k = \frac{3a - 1}{2a -2}.
\end{equation}
In equations (\ref{P_diss})-(\ref{zeta}), $\chi$
is the bulk Lorentz factor of the jet in units of its terminal  value
$\Gamma_\infty$,  
$\zeta$ (eq. \ref{zeta}) is the dimensionless 
distance in the jet
from the BH, and $\xi_\mathrm{rec}=v_\mathrm{rec}/v_\mathrm{A}$
is the fraction of the reconnection velocity relative to the local Alfvén velocity.
This reconnection rate, measured from 3D MHD numerical simulations of turbulence induced reconnection, is  $\xi_\mathrm{rec}\sim 0.03 - 0.1$  
\citep[e.g.][]{kowal_etal_09, takamoto_etal_15, singh_etal_16, delvalle_etal_16, kadowaki_etal_2021, guo_etal_2019, Medina-Torrejón_2023, Vicentin2024}.
Thus we adopt the fiducial value of
$\xi_\mathrm{rec}=0.05$ throughout the analysis of this paper (see also section \ref{sec:acctime}).

The parameter $\chi$ is obtained as solution of the differential equation:
\begin{equation}
\frac{d \chi}{d \zeta} = \frac{(1-\chi)^{k}}{\chi^2},
\label{diffeq}
\end{equation}
which has the implicit 
analytic solution:
\begin{equation}
\zeta -\zeta_0=
F(\chi) - F(\chi_0)
\label{zeta_sol}
\end{equation}
with
\begin{equation}
F(\chi)\equiv
\frac{(1-\chi)^{3-k}}{k-3}
-\frac{2(1-\chi)^{2-k}}{k-2}
+\frac{(1-\chi)^{1-k}}{k-1},
\end{equation}
being ($\zeta_0$, $\chi_0$) the initial conditions
of equation (\ref{diffeq}).

\begin{figure}
   \centering
   \includegraphics[width=\hsize]{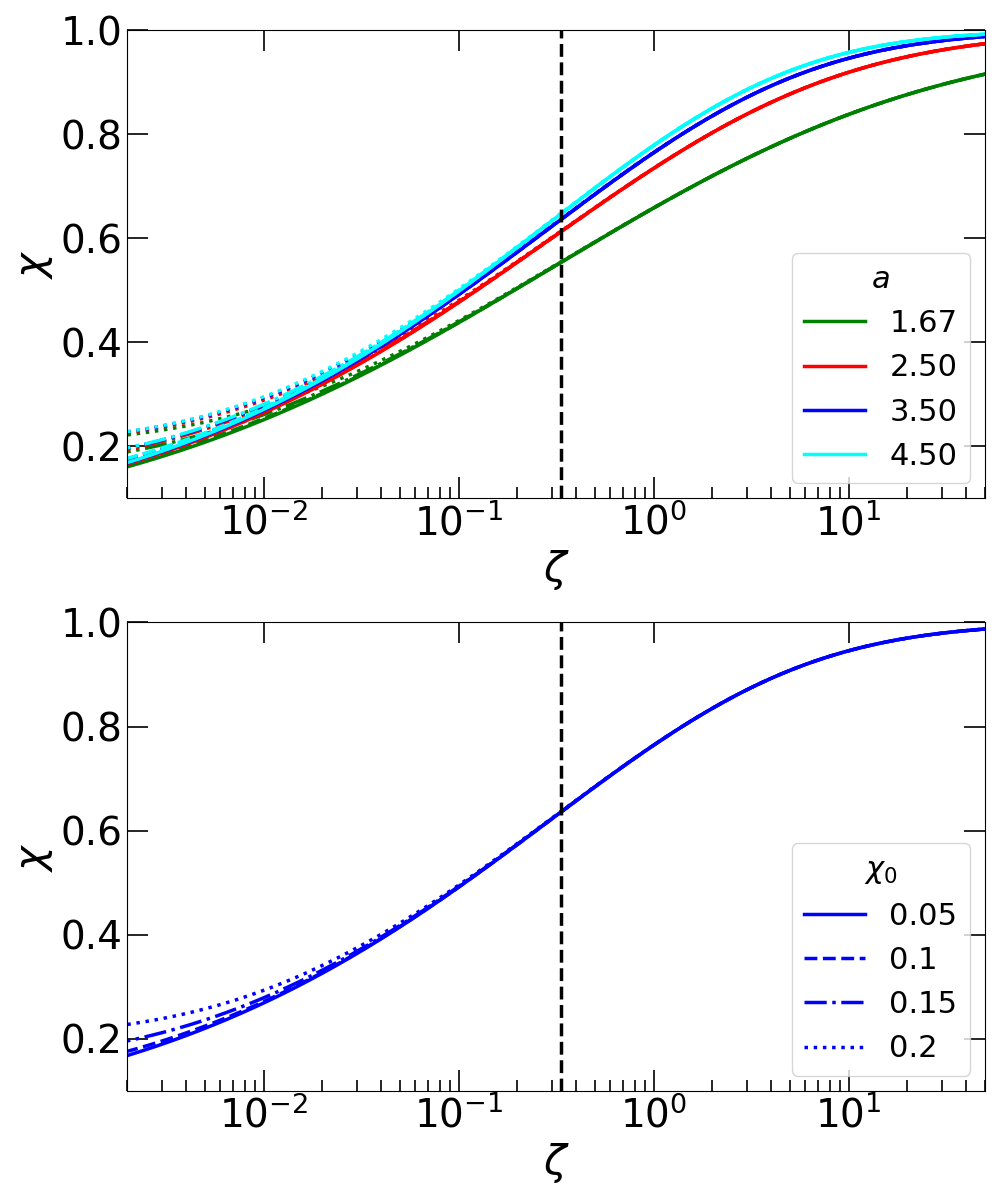}
      \caption{
Solutions to eq. (\ref{diffeq}) giving
$\chi=\Gamma_\mathrm{j}/\Gamma_\infty$
(jet bulk Lorentz factor in units of
the its terminal value)
as a function of the 
dimensionless distance to the BH $\zeta$ (defined in eq. \ref{zeta}).
Top: Curve solutions corresponding to different
values of the power-law index $a$ for the distribution of the stripes (see the text), and for different initial values of 
$\chi_0$ at $\zeta_0=0$.
Bottom: The family curve solution corresponding to $a=3.5$
is plotted alone in this panel to illustrate
the effect of different choices of $\chi_0$.
In both panels, the dashed vertical line indicates the location
where the peak of magnetic power dissipation occurs (see the text).
              }
         \label{fig:soleqdiff}
   \end{figure}

 In order to define a particular solution 
of equation (\ref{diffeq}), values
for the power index $a$
 for the distribution of the stripe widths
 and for the boundary condition
$(\zeta_0,\chi_0)$ should be defined. 
The value of the power-law index $a$
may depend on the accretion process on to the BH,
however it is 
not well constrained 
as pointed out by GU19. 
We note 
that the solutions to equation (\ref{diffeq})
are not sensitive to the choice of the value of $a$
for $a\gtrsim3.5$.
In this case the  solutions seem to converge
to an asymptotic  curve
(see Figure~\ref{fig:soleqdiff}, upper panel).
Moreover, with regard to the boundary condition,  the solutions to equation (\ref{diffeq}) 
are also practically insensitive  to 
the choice of $\chi_0$ at $\zeta_0=0$, 
for $\zeta>0.1$,  
as long as $0\leq \chi_0<<1$, as shown
in the lower panel of Figure~\ref{fig:soleqdiff}.
Thus, to simplify matters and to reduce the number of free parameters, 
here we consider the values of $\zeta_0=0$ (the location of the central engine),
$\chi_0=0.1$, and $a=3.5$ in all solutions derived in this paper.

At a distance $s$, 
the gas and magnetic energy densities in the flow co-moving frame $U'_g$ and 
$U'_\mathrm{B}$ can be related to the jet kinetic and magnetic power  as
$L_\mathrm{k} = \Gamma_\mathrm{j}^2\pi r_\mathrm{j}^2 \beta_\mathrm{j} c U'_\mathrm{g}$, and 
$L_\mathrm{B} = \Gamma_\mathrm{j}^2\pi r_\mathrm{j}^2 \beta_\mathrm{j} c U'_\mathrm{B}$, respectively, where $r_\mathrm{j}$ and $\beta_\mathrm{j}=v_\mathrm{j}/c$ are the jet cross section and flow velocity at $s$.
Then we estimate the gas density and magnetic field in the flow co-moving frame as:
\begin{equation}
\label{rho_comov}
\rho' = \frac{L_\mathrm{j}}{\Gamma_\infty^2 \pi r^2_\mathrm{j}\beta_\mathrm{j} c^3 \chi(s)},
\end{equation}
\begin{equation}
\label{B_comov}
\frac{B'^2}{4\pi} = \frac{L_\mathrm{j}\left[1-\chi(s)\right]}{\Gamma_\infty^2 \pi 
r^2_\mathrm{j}\beta_\mathrm{j} c \chi^2},
\end{equation}
where $r_\mathrm{j}\approx \theta_\mathrm{j} s$ is a good approximation for a small jet opening angle 
$\theta_\mathrm{j}$, and 
$\chi$  as a function of $s$ is obtained solving numerically equations (\ref{diffeq})-(\ref{zeta_sol}).

For calculating the physical quantities
$P_\mathrm{diss}$,
$\Gamma_\mathrm{j}$, and
${B'}$ from eqs. (\ref{P_diss})-(\ref{B_comov}), we need to define the values  of 
$\Gamma_\infty$, $L_\mathrm{j}$, and $l_\mathrm{min}$.
Given the lack of knowledge about
$\Gamma_\infty$, $L_\mathrm{j}$, and $l_\mathrm{min}$
for the source TXS 0506+056, we leave these quantities as
free parameters with the following constraints.
We take the total jet power as a factor 
$\lambda$
of the BH Eddington luminosity
$L_j=\lambda L_\mathrm{Edd}(M_{BH})$, with
$10<\lambda<100$,
as  required by hadronic models \citep[e.g.][]{2015MNRAS.450L..21Z,2020ApJ...893L..20L}.
Here we fix the mass of the central SMBH to
$M_\mathrm{BH}= 3 \times 10^{8}$ M$_\odot$ following \citet{Padovani2019}.
The minimum width of the stripes is constrained within the range of
$100R_g< l_\mathrm{min} <1000R_\mathrm{g}$, being $R_g=GM_\mathrm{BH}/c^2$ the
gravitational 
radius of the central BH, as pointed out by
GU19. 
 In addition, the terminal Lorentz factor of the jet flow
is taken within the range of $\Gamma_\infty\simeq 30-60$, which is compatible with the values typically constrained by observations of blazar jets \citep[e.g.][]{Hovatta2009, Lister2019}. 

\section{
Multi-messenger emission powered by magnetic reconnection
}
\label{sec:rad_model}

The fast magnetic reconnection driven by turbulence  among the magnetic stripes which  sustain the
stationary jet structure in the model of GU19  (Section \ref{sec:striped_model}) can
also accelerate  particles
like electrons and ions
to relativistic energies
through Fermi-like
acceleration process followed by drift acceleration
\citep[][]{dalpino_lazarian_2005, 
Medina-Torrejón_2021, Medina-Torrejón_2023}. 
In  Section \ref{sec:acctime},
we discuss these acceleration regimes in detail and show how they influence the emission of the particles and the resulting SED. We find that acceleration is dominated by the Fermi process. If these non-thermal particles carry a non-negligible fraction of the
magnetic dissipation power $P_\mathrm{diss}$ (see equation \ref{P_diss}),
their energy losses
can account for 
the observed quiescent
emission from blazars \citep[e.g.][]{Medina-Torrejón_2021, Medina-Torrejón_2023}. 

Flares or temporary enhanced jet emission could be produced by an excess of
particle injection in a localized region of the jet 
\citep[][]{aharonian_etal_07, ackermann_etal_2016, britto_elal_2016, neutrinosmm, neutrinosprior}
over the
stationary state of particle injection. 
To interpret blazar flares as an excess over its quiescent
emission, for simplicity here we calculate the emission of 
non-thermal stationary distributions of relativistic electrons
(e) and protons (p), $N_\mathrm{e,p}(E)$ (i.e., the number of particles with energy $E$ per unit volume). 
This is calculated in the frame of an approximately spherical emitting region, hereafter referred to as the blob, located in the jet's transition zone from magnetically to kinetically dominated, with a radius 
 $r_\mathrm{b}$.
We constrain the energy distributions of the relativistic particles
so that their kinetic luminosity
(in the BH frame)
represents
a fraction $\eta_\mathrm{i}$ (i = e, p) of the local magnetic dissipation by reconnection power
(see Section~\ref{sec:striped_model}):
\begin{equation}
L_i = \eta_i P_\mathrm{diss}(s) = 
\Gamma_\mathrm{j}^{2}(s) \pi r_\mathrm{b}^2 \beta_\mathrm{j} c
U'_i
\label{L_i}
\end{equation}
with $U'_i$ being the energy density of accelerated particles (electrons or protons) in the frame of the emitting reconnection blob, given as:
\begin{equation}
U'_i = \int_{E'_{0,i}}^{\infty} 
\mathrm{d} E'_i \,
E'_i N'_i(E_i),
\label{normp}
\end{equation}
where $N'_i$ is the 
number of particles per unit energy, per unit volume, in the co-moving frame of the emission region.

For relativistic protons, we approximate their stationary particle energy distribution as
\begin{equation}
N'_\mathrm{p} = 
t'_\mathrm{p}(E'_\mathrm{p})
Q'_\mathrm{p}(E'_\mathrm{p}),
\label{Np}
\end{equation}
where
$E'_\mathrm{p}$ is the proton energy,
$t'_\mathrm{p}$ is the proton injection time 
and 
$Q'_\mathrm{p}$ is the
stationary proton injection function 
(particles per unit energy, per unit volume, 
per unit time), in the blob frame.
The proton injection time is taken as
\begin{equation}
t'_\mathrm{p} = \min\left\{ t_\mathrm{p,cool}, t_\mathrm{acc} \right\},
\end{equation}
where $t'_\mathrm{p,cool}$ is the particle cooling time, and $t_\mathrm{acc}$ is the proton acceleration time, which is described in Section \ref{sec:acctime}.

The particle cooling time $t'_\mathrm{p,cool}$ is taken as the inverse of 
the total cooling rate:
\begin{equation}
\label{t_pcool}
t'_\mathrm{p,cool} =
\left[
t'^{-1}_\mathrm{p\gamma}(E'_\mathrm{p}) +
t'^{-1}_{\mathrm{B-H}}(E'_\mathrm{p}) +
t'^{-1}_\mathrm{p,syn}(E'_\mathrm{p}) 
\right]^{-1},
\end{equation}
being,
$t'^{-1}_\mathrm{p\gamma}$, 
$t'^{-1}_\mathrm{B-H}$, and 
$t'^{-1}_\mathrm{p,syn}$
the proton cooling rates due to photo-pion 
production, Bethe-Heitler pair production, and
synchrotron radiation, respectively\footnote{
We note that the cooling of CR protons due to proton-proton reactions is generally not relevant in blazar jets and specially in this system due to the very low density of the thermal plasma.}.  
We refer to  Appendix~\ref{sec:Appendix A} for the mathematical
expressions that we employ for the cooling 
rates of equation (\ref{t_pcool}).

We parametrise the proton injection function $Q'_\mathrm{p}$
(particles per unit energy, per unit volume, per unit time)
in eq. (\ref{Np}) as a stationary power-law energy distribution with squared exponential cut-off:
\begin{equation}
Q'_\mathrm{p} = 
Q'_0(E'_\mathrm{p}/E'_\mathrm{p,0})^{-\alpha_\mathrm{p}}
\exp\left\{
- \left(E'_\mathrm{p} / E'_\mathrm{p,max}\right)^{2}
\right\},
\label{Q_p}
\end{equation}
where we set $E'_\mathrm{p,0}= 1$ GeV as the minimum energy of the proton distribution, 
$\alpha_\mathrm{p}$ is the proton power-law index,
$E'_\mathrm{p,max}$ is the  maximum energy of the proton
population (see Section \ref{sec:acctime}), and the normalisation factor $Q'_0$ is obtained through equations (\ref{L_i}-\ref{normp}).

For the parameter space explored in this work, 
the SSC cooling  of primary electrons mostly falls in the Thompson regime, 
and the cooling time scale due to synchrotron and SSC radiation of the accelerated electrons is always much shorter than the adiabatic time scale (Section~\ref{sec:SED}).
Thus, we parametrise the stationary energy distribution of primary electrons (in units of particles per volume per energy) as a single power-law with squared exponential 
cutoff:
\begin{equation}
N'_\mathrm{e}  =
N'_\mathrm{e,0}
\left(
E'_\mathrm{e}/E'_\mathrm{e,0}
\right)^{-n_\mathrm{e}}
\exp\
\left\{
-\left(E'_\mathrm{e}/E'_\mathrm{e,max}\right)^2
\right\},
\label{Ne_bpl}
\end{equation}
where $E'_\mathrm{e,0}$ 
and $E'_\mathrm{e,max}$
are the minimum and maximum
energy of the electron population, respectively, and
we set $n_\mathrm{e}=\alpha_\mathrm{p}+1$.
This choice for the index $n_\mathrm{e}$ is motivated by a scenario where
electrons and protons are energised by the same acceleration process.
Finally, the normalisation factor $N'_\mathrm{e,0}$ of the 
electron distribution 
is also obtained through eq. \ref{normp}.





\section{Model for the TXS 0506+056 MM flare}

\label{sec:app_to_txs}
As stressed, here we interpret the MM SED data set related to  the 2017 neutrino flare
of TXS 0506+056 \citep[][]{neutrinosmm}, 
assuming a lepto-hadronic emission powered by magnetic reconnection
in the blazar jet (see Sections~\ref{sec:striped_model} and
\ref{sec:rad_model}).
The broadband SED of the source associated with the detection of the neutrino event IC-170922A is shown in Figure~\ref{fig:SEDs_M3e8}, adapted from \cite{neutrinosmm}.
The fluxes indicated by the upper limits at ~[0.3–3] PeV, with solid and dashed lines, are consistent with one neutrino detection within 0.5 years and 7.5 years, respectively. The data points from radio to VHE gamma-rays present the quasi-simultaneous SED observed from the source. The radio data points, plotted in grey, are not intended to be explained by the emission scenario discussed here, as this emission in blazars is generally attributed to a much larger-scale region in the jet.

We particularly work out the hybrid radiation scenario where  
CR protons account for the production of HE neutrinos
in which the high-energy EM bump of the blazar spectrum 
is mostly accounted by Synchrotron-self-Compton (SSC)  emission of primary electrons.
Under this radiation scenario, we expect 
partial contributions from CR protons to the EM
emission at X-rays and VHE gamma rays through electromagnetic cascades, 
as exhibited by hybrid models \citep{Cerruti_2019, Gao_2019}.


\subsection{The location of the emission region} \label{location}

We combine the aforementioned SSC radiative condition with the
jet model described in Section~\ref{sec:striped_model} to 
constrain the location range in the jet where the conditions are fulfilled to produce the observed MM emission. 
Requiring that the bumps of the blazar spectrum peaking at $E_\mathrm{pk,\ell}$ and $E_\mathrm{pk,h}$
are produced mostly by synchrotron and SSC radiation, respectively, we first constrain the possible values of $B'$ and $\Gamma$ of the emission region as follows.

We note that in the co-moving frame of the emission region, relativistic electrons with Lorenz factor $\gamma_\mathrm{e}'$ produce syncrhotron emission with frequency
$\nu_\mathrm{syn}'=\gamma_\mathrm{e}'^2\nu_\mathrm{L}'$, being $\nu'_\mathrm{L}=e B'/(2\pi m_\mathrm{e}c)$ the Larmor frequency, and $e$ the electron charge \citep{Longair_2011}.
The associated SSC photons boosted by the same electrons emit at frequency
$\nu'_\mathrm{ssc}\approx\gamma_\mathrm{e}'^2\nu'_\mathrm{syn}$ 
 \citep{Ghisellini_2013}.
Thus, the frequencies of synchortron and SSC peaks can be related as
$\nu_\mathrm{syn}'\approx(\nu_\mathrm{ssc}'/\nu_\mathrm{syn}')\nu_\mathrm{L}'$. On the other hand,
the synchrotron and SSC frequencies in the observer frame can be approximated as $\nu_\mathrm{syn}=\nu_\mathrm{syn}'\Gamma_\mathrm{j}/(1+z)$ and 
$\nu_\mathrm{ssc}=\nu_\mathrm{ssc}'\Gamma_\mathrm{j}/(1+z)$, respectively. Therefore, given the energies of  the low and high energy bumps of the blazar spectrum,
the co-moving magnetic field and the jet bulk Lorentz factor at the emission region can be constrained as:
\begin{equation}
B'=
\frac{E_\mathrm{pk,\ell}^2}{E_\mathrm{pk,h}}
\frac{m_\mathrm{e} c}{e\hbar}
\frac{(1+z)}{\Gamma_\mathrm{j}}.
\label{tilB_G_SSC}
\end{equation}
In this equation, $E_\mathrm{pk,\ell}$ and $E_\mathrm{pk,h}$ are the   low and high energy  blazar SED bumps, respectively, at the observer's frame.
We note that SED models with energies values in the interval of 
$E_\mathrm{pk,\ell}\in[5, 15]$eV are consistent with  observed low energy blazar bumps  and, at the same time, have suitable target photons to produce high energy neutrinos peaking at $\sim$ PeV energies (in the observer's frame).
On the other hand,  the observed high energy bump  of TXS 0506+056  is consistent with $E_\mathrm{pk,h}\sim [0.5- 2]$ GeV  interval \citep[][]{neutrinosmm}.

In order to find jet solutions and derive the location of the emission region ($s_\mathrm{em}$) which are
consistent with the constraint given by eq. \ref{tilB_G_SSC}, we employ
 equations \ref{zeta_sol} to \ref{B_comov}. This combination results:

\begin{align}
\label{sem}
s_\mathrm{em} &= \frac{l_\mathrm{min}\Gamma_\infty^2}{2\xi_\mathrm{rec}}
\left[
F(\chi_\mathrm{em}) - F(\chi_0)
\right],\\
\label{xem}
\chi_\mathrm{em} &= 1 -
\frac{c\beta_\mathrm{em}}{L_\mathrm{j}}
\left[
\frac{E_\mathrm{pk,\ell}^2}{E_\mathrm{pk,h}}
\frac{m_\mathrm{e}c(1+z)}{2e\hbar}\theta_\mathrm{j}s_\mathrm{em}
\right]^2,
\end{align}
where  $\beta_\mathrm{em}=\left(1-(\Gamma_\infty\chi_\mathrm{em})^{-2}\right)^{1/2}$.

\begin{figure}
\centering
\includegraphics[width=\hsize]{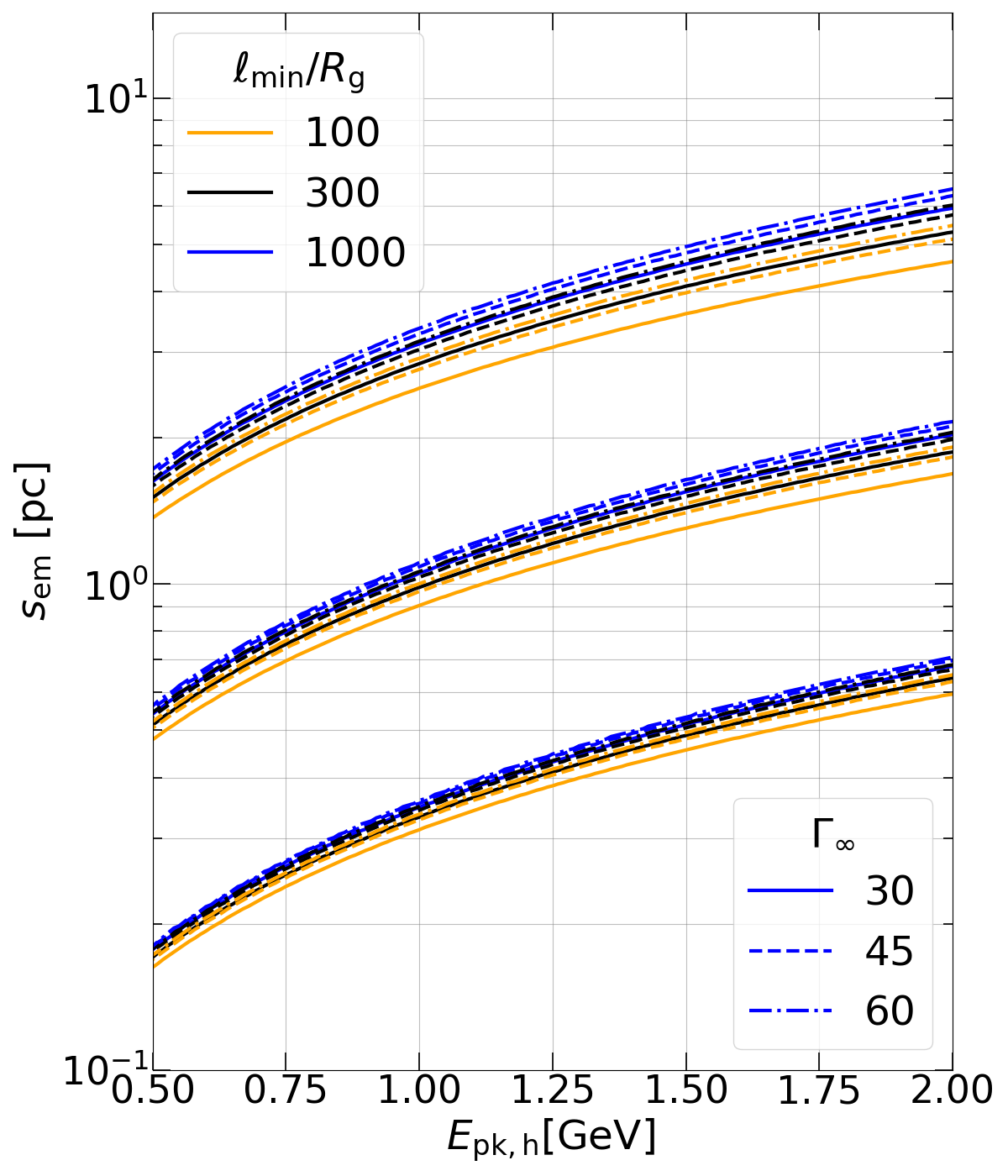}
\caption{
Constraining the location of the emission region along the jet.
We display the solutions to equations
\ref{sem}-\ref{xem}, for the possible location
$s_\mathrm{em}$.
Curves plotted with different colours and line styles  correspond to different values of 
minimum length of the stripes $l_\mathrm{min}$ and of the terminal Lorentz factor $\Gamma_\infty$ of the jet flow,
as indicated. 
The lower, middle, and upper groups of curves correspond to total jet powers of $L_\mathrm{j}=1.5, 15$ and 150$L_\mathrm{Edd}$, respectively.
}
\label{fig:s_em}
\end{figure}

In Figure \ref{fig:s_em}, we display the solutions for the  emission region, $s_\mathrm{em}$,
as a function of the observed energy range  of the high bump of the TXS 0506+056 SED within
the interval $E_\mathrm{pk,h}\in[0.5,2]$ GeV, fixing $E_\mathrm{pk,\ell}=10$  eV.
The different colours and line styles of the 
curves 
correspond to different values of $l_\mathrm{min}/R_\mathrm{g}$ ($100, 300$, and $1000$),
and $\Gamma_\mathrm{\infty}$ ($30, 45$, and 60), as indicated. They are calculated for three  values of the jet power  $L_\mathrm{j}=1.5, 15$ and 150$L_\mathrm{Edd}$.

\begin{figure*}
   \centering
   \includegraphics[width=0.9\hsize]{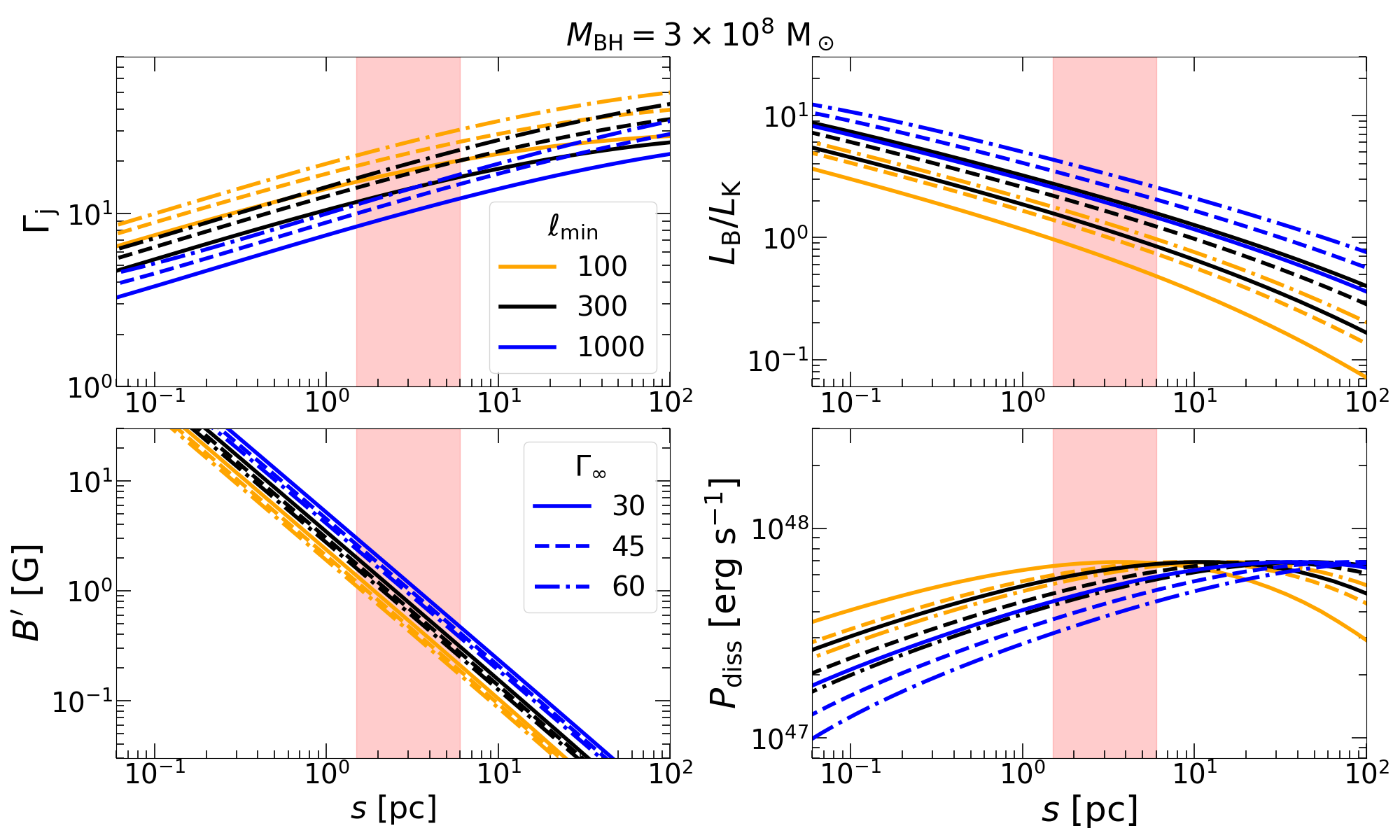}
      \caption{
Jet properties given by the stripe jet model  
(Section~\ref{sec:striped_model}) as functions of
the distance from the SMBH. All curves in this plot are
calculated assuming an SMBH of $M_{BH}  =3\times10^{8}$ M$_\odot$,
and a jet total power of  $L_\mathrm{j} =$ 
150$L_\mathrm{Edd}(M_\mathrm{BH})$.
The pink shaded regions indicate the location interval
compatible with the constraint obtained 
from Figure~\ref{fig:s_em} for the possible emission regions
(see the text).
The different curve styles in this plot are calculated assuming
different values for  the terminal Lorentz factor of the jet flow
as labelled.
Orange, black  and blue colours indicate curves obtained assuming
$l_\mathrm{min}/R_\mathrm{g}=$ 100, 300, and 1000, respectively
($R_\mathrm{g} = GM_{BH}/c^2$). These intervals will serve as guidelines for modelling the MM SED of TXS 0506+056. 
}
         \label{fig:strip_sols}
   \end{figure*}

Considering  $L_\mathrm{j}=150  L_\mathrm{Edd}$ 
from Figure~\ref{fig:s_em}, 
we see that the possible locations for the emission regions
can be limited within the interval of $\Delta s_\mathrm{em}\sim$[1.5 - 6] pc. 
For this length range,  the jet dissipation power 
$P_\mathrm{diss}$,
magnetization 
$\sigma = L_\mathrm{B}/L_\mathrm{K}$,
magnetic field 
$B'$, and
$\Gamma_\mathrm{j}$, 
fall within the range of values indicated by the pink shaded regions in Figure \ref{fig:strip_sols}.

\begin{figure}
   \centering
   \includegraphics[width=0.9\hsize]{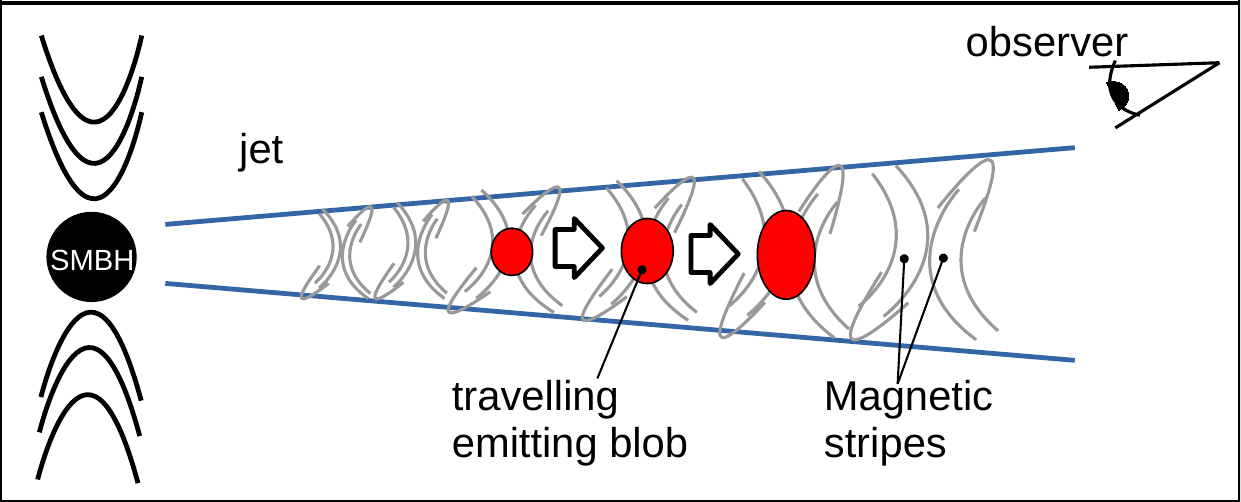}
      \caption{ 
Sketch of the emitting blob as it moves downstream the expanding jet. The three different positions indicate the regions at which the SED of TXS 0506+056 is evaluated in Figure 
\ref{fig:SEDs_M3e8}. The blob is assumed to be spherical in the comoving frame. 
}
         \label{fig:sketch}
   \end{figure}

Based on the parametric space defined above, 
we  construct the lepto-hadronic SED  for TXS 0506+056. To achieve this we   consider the emitting blob, described in section \ref{sec:rad_model}, moving downstream with the jet flow at three different positions, which are illustrated in Figure \ref{fig:sketch}. 
At each position along the jet,  the blob radius is constrained by the causality condition given by the minimum variability time interval $\Delta t_\mathrm{v}$ observed during the 2017 flare. This results
\begin{equation}
r_\mathrm{b} = f_\mathrm{v}\frac{c\Delta t_\mathrm{v}\Gamma_\mathrm{j}}{(1+z)},
\end{equation}
where $f_\mathrm{v}$ is a dimensionless  free parameter  $f_\mathrm{v}\leq1$, and
we take $\Delta t_\mathrm{v} \simeq 1$ day, which is consistent with the minimum variability of the source observed at X-rays  \citep{Keivani_2018}.

\subsection{Particle acceleration  by magnetic reconnection and the maximum energy of the particles }

\label{sec:acctime}



As remarked in Section \ref{sec:intro}, during the Fermi regime 
of particle acceleration through reconnection, it is expected that particles  experience an exponential growth in energy over time \citep[e.g.][]{dalpino_lazarian_2005}. 3D MHD simulations with test particles align with this expectation, revealing a reconnection acceleration time with a weak dependence on particle energy, expressed as $t_{acc} \propto E^{-0.2-0.1}$, specially for relativistic reconnection velocities \citep[e.g.][]{delvalle_etal_16, 2017PhRvL.118h5101L,Medina-Torrejón_2021, Medina-Torrejón_2023}.
In a recent study, \citet{xu_lazarian_2023}  revisited the earlier work by \citet{dalpino_lazarian_2005}, and derived the following condition for  fastest acceleration time (in the Fermi regime) within a turbulence-induced magnetic reconnection layer of reconnection velocity $v_\mathrm{rec}$ and  thickness $\Delta$:
\begin{equation}
  t_\mathrm{acc} \sim \frac{4 \Delta}{c d_\mathrm{ur}}  
  \label{tacc3}
\end{equation}
where 
\begin{equation}
d_\mathrm{ur} \approx \frac{2\beta_\mathrm{rec}(3\beta^2_\mathrm{rec}+3\beta_\mathrm{rec}+1)}{3(\beta_\mathrm{rec}+ 0.5)(1-\beta^2_\mathrm{rec})},
\end{equation}
with $\beta_\mathrm{rec} = v_\mathrm{rec}/c$ and $c$  the light speed. 
This relation has been recently successfully confirmed by means of 3D MHD simulations of turbulent relativistic jets with test particles \citep[][]{medina-torrejon_dalpino2024}.

In this work $\Delta$ corresponds to  the size of the acceleration region which in turn is given by the blob  
size $\Delta \sim  2 r_\mathrm{b}$. 
We take from the numerical simulations   $v_\mathrm{rec} \simeq 0.05 v_\mathrm{A}$ \citep[e.g.][]{Medina-Torrejón_2021, Medina-Torrejón_2023}, where the Alfvén speed in the relativistic regime is:
 \begin{equation}
v_\mathrm{A} \approx \frac{v_\mathrm{A,0} }{\sqrt{ 1 + \left(\frac{v_\mathrm{A,0}}{c}\right)^2}}
\end{equation}
where $v_\mathrm{A,0} = B'/\sqrt{4 \pi \rho'}$. 
For the jet model adopted in this work, $v_\mathrm{A,0}$ can be calculated through equations (\ref{rho_comov}) and (\ref{B_comov}), giving:
\begin{equation}
v_\mathrm{A,0} = c\sqrt{\frac{1-\chi}{\chi}},
\label{vA0}
\end{equation}
where $\chi$ depends on the jet location as given by equations (\ref{diffeq}) -(\ref{zeta_sol}).

This  acceleration time is independent of the particles energy, as expected in the Fermi regime for constant reconnection speed, and which is in agreement with  the above mentioned simulations  \citep[see also][]{medina-torrejon_dalpino2024}.

\begin{figure}
   \centering
   \includegraphics[width=\hsize]{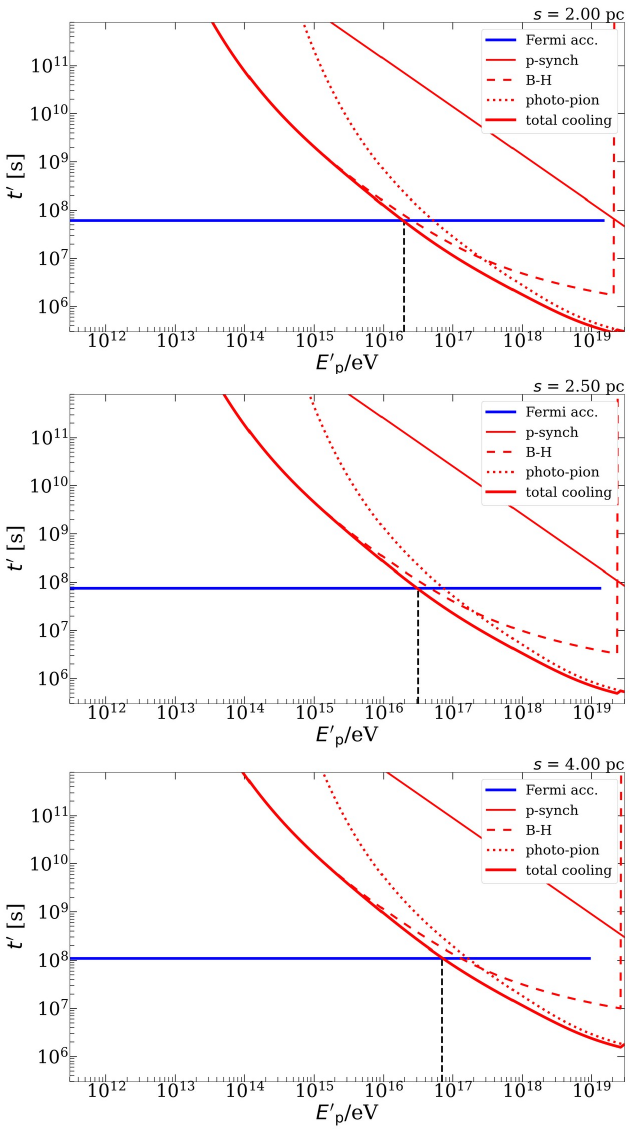} 
      \caption{
      Cooling time of the CR protons (source frame) which  produce the hadronic emission components of the SED spectrum  in Figure~\ref{fig:SEDs_M3e8}, calculated for three different positions of the emission blob in the jet, $s=$ 5 (top), 8 (middle) and 10 pc (bottom panel).
The different cooling processes (photo-pion production, photo-pair (B-H) production, and  synchrotron) are displayed with thin curves of different line styles as indicated, whereas the total cooling time is plotted by the red thick curve.
 We over-plot the proton acceleration time due to magnetic reconnection in the Fermi acceleration regime (blue line). The resulting total radiative cooling time curve is intercepted by the acceleration time curve at the vertical dashed black line. This 
indicates the maximum energy required in our modelling to produce the theoretical SED sequence of Figure~\ref{fig:SEDs_M3e8} (see text for details). 
}
 \label{fig:coolingHAD_M3e8}
   \end{figure}

This regime of acceleration will persist until a threshold energy which is attained when the particles Larmor radius becomes larger than the thickness of the reconnection layer, 
$r_\mathrm{L}  \simeq  2 r_\mathrm{b}$.
For instance, for the parameters of the inner blob position, this implies a proton energy threshold $E_\mathrm{th}
\simeq 10^{19} $ eV (Figure \ref{fig:coolingHAD_M3e8}).
 This value is also consistent with those obtained from 3D MHD numerical simulations of relativistic jets with test particles for similar background magnetic fields, in the absence of radiative losses \citep[][]{Medina-Torrejón_2021}.


In a subsequent drift regime, occurring after particles have attained the threshold energy and exited the reconnection zones, the energy growth with time becomes strongly dependent on the energy and therefore, slower. The acceleration time in this regime is approximately described by the equation $t_\mathrm{acc,drift} \simeq E_\mathrm{p}/(q B' v_\mathrm{rec})$ 
\citep[][]{dalpino_kowal_15, delvalle_etal_16, Zhang2023}.
The MHD numerical simulations with test particles have confirmed that the extended acceleration time observed in the drift regime is attained only for $E_p > E_{th}$  \citep[][]{kowal_etal_2012, delvalle_etal_16, Medina-Torrejón_2021}.


\begin{figure}
   \centering
   \includegraphics[width=\hsize]{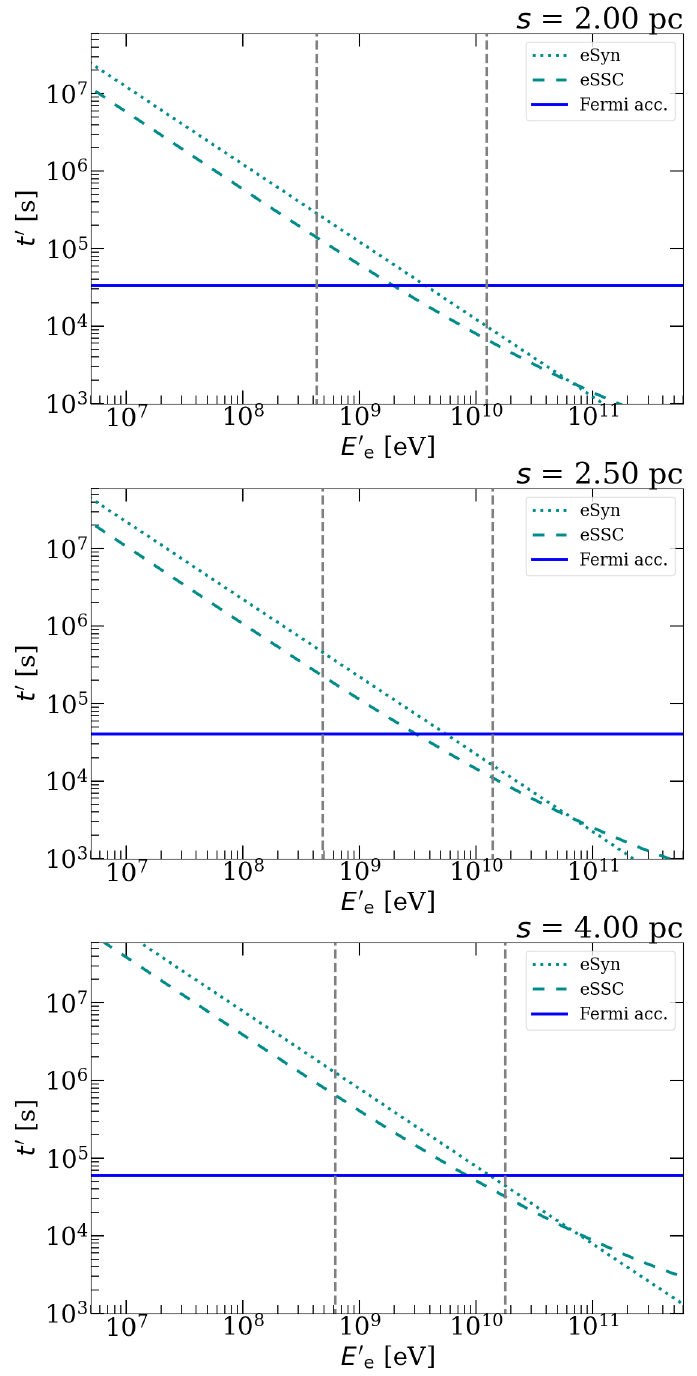} 
      \caption{
Cooling times (green lines) and reconnection acceleration time in the Fermi regime (blue line)  of accelerated electrons in the frame of the emitting blob for the SED sequence presented in Figure~\ref{fig:SEDs_M3e8}. 
The vertical dashed black lines indicate the minimum (left)
and the maximum energy (right) of the electron population evaluated directly from the synchrotron peak around  3 eV of the SED of TXS 0506+056).    
}
\label{fig:coolingLEP_M3e8}
   \end{figure}

In Figure \ref{fig:coolingHAD_M3e8}, 
 we compare the particle acceleration time  by magnetic reconnection 
with  the total proton radiative cooling time  (eq. \ref{t_pcool}). 
both calculated  in the plasma frame, in three different positions along the jet. These positions correspond to those at which the SED is  calculated in Figure \ref{fig:SEDs_M3e8} (Section \ref{sec:SED}). 


The
maximum possible energy of the protons in the emission blob, $E'_\mathrm{p,max}$  (eq. \ref{Q_p}),
in Figure \ref{fig:coolingHAD_M3e8},
 is given by the value at which  $$ t'_\mathrm{acc} = t'_\mathrm{cool}.$$

We observe  from the very high energy threshold  found  for the Fermi regime in the three diagrams of Figure 
\ref{fig:coolingHAD_M3e8}
 (indicated by the maximum extension of the blue lines) that this regime will  dominate particle acceleration within the blob. 
We also note that the obtained values of $E'_\mathrm{p,max}$ are well below this  threshold energy ($E_{th}$),  
and protons with energies 
$E_\mathrm{p} > E'_\mathrm{p,max}$
 will  lose their energy radiatively inside the emission blob.

Similarly, the maximum energy for the electrons,
$E'_\mathrm{e,max}$ (eq. \ref{Ne_bpl}), can be also
 obtained from the balance between the total radiative cooling time and the reconnection acceleration time.
Nevertheless, it is worthy noting that the supra-thermal electrons in the acceleration region  will start Fermi acceleration at scales much smaller than that of  the protons.  This is due to their much smaller Larmor radius for energies of the order of their rest mass. They will start interacting with very tiny magnetic fluctuations at the acceleration zone, effectively in reconnection structures with thickness $\gtrsim r_{L,e}$. At these scales, $\Delta$ in eq. \ref{tacc3} results an acceleration time for the electrons which is much smaller than that  evaluated for protons  in Figure \ref{fig:coolingHAD_M3e8}. From the ratio between the electron and the proton momentum equations in the relativistic regime, one can easily demonstrate that $t'_\mathrm{acc,e}/t'_\mathrm{acc,p} \sim m_e/m_p$ \citep[see also][]{khiali_etal_15, Khiali2016, Medina-Torrejón_2021}.



Figure \ref{fig:coolingLEP_M3e8} presents the estimated Fermi acceleration time, as calculated from Eq. \ref{tacc3} and adjusted by the mass ratio factor, alongside the synchrotron and SSC loss times for electrons at the three different locations of the emitting blob in the jet. The vertical grey dashed line on the right denotes the maximum electron energy, \(E_{e,\text{max}}\), calculated to match the observed synchrotron bump in the SED of TXS 0506+056.
Notably, the reconnection acceleration time intersects the synchrotron cooling time curves near this maximum energy, indicating that the acceleration model aligns well with the maximum electron energy inferred from observations.

\begin{table} 
	\centering
	\caption{Free  and calculated  parameters  from the jet emission model at each position $s_\mathrm{em,i}$ of the blob, leading to the  SED curves of 
 Figure \ref{fig:SEDs_M3e8}.
Quantities in blue
with the (*)  label 
are the derived parameters from model. }
	\label{tab:example_table}
	\begin{tabular}{lccr} 
		\hline
		Parameter & s$_\mathrm{em,1}$& s$_\mathrm{em,2}$ & s$_\mathrm{em,3}$ \\
		\hline
		$\lambda = L_\mathrm{j}/L_\mathrm{Edd}$ & 1.5E2 & 1.5E2 & 1.5E2 \\
		  $f_\ell=\ell_\mathrm{min}/R_\mathrm{g}$ & 1.0E3 & 1.0E3 & 1.0E3\\
		  $\Gamma_\infty$ & 45 & 45 & 45 \\
		  $s$ [pc] & 2.0 & 2.5 & 4.0 \\    
        $f_\mathrm{v}$  & 0.672 & 0.753 & 0.948  \\
		  $\alpha_\mathrm{p}$ & 1.7 & 1.7 & 1.7 \\
		  $\eta_\mathrm{p}$ & 0.75 & 0.65 & 0.50 \\
        $\eta_\mathrm{e}$  & 2.2E-4 & 2.5E-4 & 3.2E-4 \\
        $E'_\mathrm{e,0}$ [MeV] & 4.32E2 & 4.85E2 & 6.22E2 \\
        $E'_\mathrm{e,max}$ [GeV]  & 1.25E1 & 1.40E1 & 1.80E1\\
		  \textcolor{blue}{$L_\mathrm{j}$ [erg/s] *} & 5.67E48 & 5.67E48 & 5.67E48 \\           
        \textcolor{blue}{$\chi (s)$ * } & 0.24 & 0.26 & 0.29 \\  
        \textcolor{blue}{$\Gamma_\mathrm{j}(s)$ * } & 10.8 & 11.6 & 13.2 \\ 
		  \textcolor{blue}{$B'(s)$ [G] *} & 1.79 & 1.33   & 0.71 \\    
        \textcolor{blue}{$L_\mathrm{B}/L_\mathrm{K}(s)$ * } & 3.15 & 2.90 & 2.41 \\
        \textcolor{blue}{$v_\mathrm{rec}/c$ * } & 4.36E-2 & 4.31E-2 & 4.20E-2 \\    
        \textcolor{blue}{$r_\mathrm{b}=f_\mathrm{v}\frac{c \Delta t_\mathrm{v}\Gamma_\mathrm{j}(s)}{(1+z)}$ [cm] * } & 1.41E16 & 1.69E16 & 2.43E16 \\    
		  \textcolor{blue}{$E'_\mathrm{p,max}$ [PeV] *} & 1.98E1 & 3.14E1 & 5.58E1 \\   
  \hline
	\end{tabular} \label{table1}
\end{table}


\subsection{Multi-messenger SEDs} \label{sec:SED}

\begin{figure}
   \centering
   \includegraphics[width=\hsize]{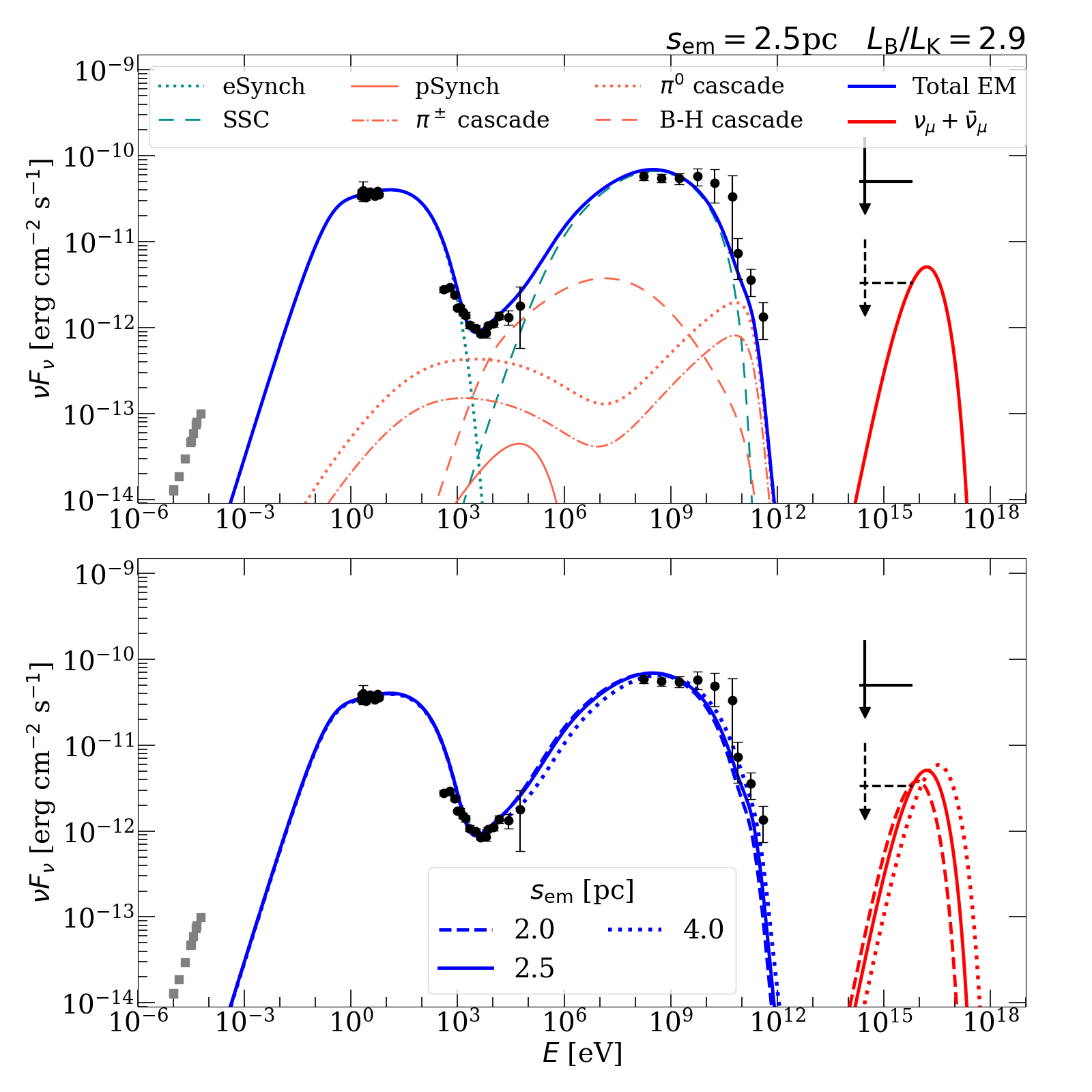} 
      \caption{
Blazar MM spectrum profiles
computed based on
lepto-hadronic emission powered by magnetic
reconnection in the jet.
The curves in this Figure are calculated by 
the jet model described in Section~\ref{sec:striped_model}
together with the lepto-hadronic radiation 
model detailed in Section~\ref{sec:rad_model}.
Top: EM flux (blue  curve) and the associated single-flavor neutrino flux (muon plus anti-muon neutrinos, red curve) when the emitting blob is located at 2.5 pc from the SMBH. This panel illustrates the radiative components of leptonic and hadronic origin, as indicated.  
Bottom: total EM and muon neutrino emission, plotted in blue and red, respectively, at different locations $s$ in the blazar jet, as labeled.  
The values of the input and output parameters for these emission profiles are reported in Table~\ref{table1}.
The over-plotted data points correspond to the 
observed MM SED of the 2017 neutrino flare from the 
source TXS 0506+056, adapted from \citet{neutrinosmm}.
}
\label{fig:SEDs_M3e8}
   \end{figure}

We derive lepto-hadronic SED profiles from the emission region, which is a spherical blob moving downstream in the jet (Figure \ref{fig:sketch}).
This region has uniform macroscopic properties, parametrized according to the GU19 model, as described in Section~\ref{sec:striped_model}.
The properties of the blob are then calculated as function
of the jet longitudinal distance $s$.
An example of the calculated MM SED at a fixed location in the jet is shown in the upper panel of Figure~\ref{fig:SEDs_M3e8}, where we display the different emission components calculated as described in Section~\ref{location} and Appendix~\ref{sec:Appendix B}.  
In the lower panel of Figure~\ref{fig:SEDs_M3e8}, we show the outcome of this model, probing different locations as the blob moves downstream within the interval $\Delta s \rightarrow [2-4]$ pc.

This is within 
the  range $\Delta s_\mathrm{em}$
constrained in  
Figure \ref{fig:s_em} and Figure \ref{fig:strip_sols} (pink band).


Each SED of the sequence shares the same values for the parameters $L_{j}$, $\Gamma_{\infty}$, and $l_\mathrm{min}$.
Then, each SED 
is obtained
by freely tuning the other parameters 
$f_\mathrm{v}$,   
$\eta_\mathrm{e}$, 
$\eta_\mathrm{p}$,
$\alpha_\mathrm{p}$,
$E'_\mathrm{e,0}$,
and $E'_\mathrm{e,max}$ 
to match the TXS 0506+056 SED data set (see Sections~\ref{sec:rad_model} and \ref{sec:acctime}). The assumed and derived   parameters of the emitting blob at the different positions along the jet are given in Table \ref{table1}. 

The SED with the peak  of the neutrino flux closest to the 
Icecube upper limits  
is obtained at the blob location
closest  to the jet core ($s=2$ pc),
corresponding to a region of higher magnetization
$\sigma\equiv L_\mathrm{B}/L_\mathrm{K}$ (bottom panel).
This 
SED 
does not reproduce yet the VHE gamma-rays (> 100 GeV) of the data set, only the high state peak around 1 GeV, as expected from the observations.


As 
the blob moves  to  larger distances $s$,
the peak of the neutrino flux moves  towards higher energies, and hence outside the sensitivity energy interval  of the Icecube detector, while the flux of VHE $\gamma$-rays
(E>100 GeV) is enhanced 
and finally, when the blob moves 
to $s=4$ pc,
the largest position within the considered range, 
the  SED matches  the observed EM data (including
the VHE gamma-rays), but exhibits the lowest flux of neutrinos.

The behaviour of this SED sequence
is consistent with the arrival time delay between the HE neutrino 
observed in the direction of 
TXS 0506+056 (which is simultaneous to the   high state  emission of the source around $\sim$1 GeV), 
and the subsequent
appearance of the VHE $\gamma$-ray signal, followed by the loss of neutrino detection. 

The time interval comprising the sequence of SEDs shown in the lower panel of 
Figure \ref{fig:SEDs_M3e8}
 can be estimated as follows. As the blob moves downstream, it suffers relativistic  beaming with apparent superluminal motion, that must be corrected according to \citep[e.g.][and references therein]{Rees1966, dalpino1991}

\begin{equation}
\Delta t_\mathrm{ap} = \Delta t (1+z) 
[1-\bar{\beta}_\mathrm{j} \cos(\theta_\mathrm{j})]
\approx
\Delta t \frac{(1+z)}{\bar{\Gamma}_\mathrm{j}^2}
\label{sluminal}
\end{equation}
where the last approximation is obtained considering a second order 
expansion of $\cos(\theta_\mathrm{j})$
and 
$\theta_\mathrm{j}\sim 1/\bar\Gamma_\mathrm{j}$.
In eq. \ref{sluminal}, $\Delta t$ is the
time interval between the emission sequence within the length interval $\Delta s$,
as measured in the BH frame.
We 
evaluate this time interval as $\Delta t = \Delta s/\bar{v}_j\sim \Delta s/c$,
being $\bar{v}_\mathrm{j}=c \bar{\beta}_\mathrm{j}$ the average velocity of the jet flow
within $\Delta s$.
Using $\Delta s = 0.5$pc, $\bar{\Gamma}_\mathrm{j}\sim 11$, and  $z\simeq 0.34$, 
eq. \ref{sluminal} gives an observed time interval of $\Delta t_{ap} \sim 6.6$ days between the first and the second SEDs of  Figure~\ref{fig:SEDs_M3e8} (bottom). This is consistent with the observed time elapsed between the neutrino and the VHE energy flare ($>100$ GeV) of    
TXS 0506+056 2017 \citep[][]{neutrinosmm}.
Likewise, the total  time interval between the first and the last  SED in  Figure~\ref{fig:SEDs_M3e8} (bottom), 
$\Delta t_{ap} \sim 26.4$ days, is  compatible  with an observed persistence of the VHE emission and the  disappearance of the neutrino signal in the 2017 event.  

\section{Summary and discussion}
\label{sec:summary}

In this paper, we  model the multi-messenger association of the
2017 neutrino flare from the blazar TXS 0506+056 as due to magnetic dissipation in the blazar jet, and derive the MM emission  as driven by magnetic reconnection. 
We adopt a single-zone lepto-hadronic model, where the emission region is represented by a spherical blob moving downstream along with the expanding jet flow, producing the expected observational signals at multiple locations.

The SEDs derived  are based on the
following considerations:

\begin{enumerate}

\item 
The mass of the central BH is
$\sim 3 \times 10^{8}$ M$_\odot$;

\item
The moving emission blob is located at the jet's transition region from magnetically to kinetically dominated, where particle acceleration by magnetic  reconnection is expected to 
 be  
 the primary mechanism;
\item
There is  no influence from  external low-energy photons on the emission
volume and hence the synchrotron photons of the accelerated   electrons internally 
provide the target radiation field for inverse Compton (IC) scattering, photo-hadronic interactions, and $\gamma-$ray attenuation within the emission volume; 

\item
The high energy peak ($E_{pk}\sim 1 $ Gev) of the blazar spectrum is dominated by the 
SSC emission of the primary electrons;

\item
The macroscopic properties of the emission volume, such as the local magnetic field, 
the jet bulk Lorentz factor, and the magnetic dissipation power are parametrised 
as  functions of the jet propagation axis $s$, following the model of GU19.

\end{enumerate}

Combining the conditions (i),  (iv) and (v), we have constrained
the possible locations of the flare emission in the jet within
the range $\Delta s\sim$ [2- 4] 
pc,
launching a jet with total 
power of $L_\mathrm{j} = 150 L_\mathrm{Edd}(M_\mathrm{BH}$) 
(see Figures \ref{fig:s_em}
and \ref{fig:strip_sols}). 
We then calculated MM SED profiles with the lepto-hadronic model
at multiple  positions along  the range $\Delta s$.
Comparing  the calculated SED models to the SED data set of
the 2017 MM flare from TXS 0506+056, we find 
the following features.

The most intense observable flux of HE neutrinos is produced at the     
position, within the range $\Delta s$, which is closest to the jet core, where magnetisation is 
highest, as expected.
At this position, the SED profile also  reproduces the HE state component ($\sim 1$ GeV)  
of the SED data set. As the emitting blob moves further downstream,  the calculated SED is consistent with the observed  electromagnetic  data  
including the VHE gamma rays ($E > 100$ GeV)
and the farthest position within 
$\Delta s$  has 
the lowest flux of neutrinos (out of the reach of the IceCube instrument). 
The behaviour of this sequence of 
emission profiles is consistent with the observed time delay between
the arrival of the HE neutrino from the direction of 
TXS 0506+056 (which is simultaneous with the high state 
$\gamma$-ray emission at  $\sim$1 GeV), 
and the subsequent
appearance of the VHE signal.

The time interval corresponding to the covered range $\Delta s$
of the SED sequence must be  corrected for the apparent superluminal motion that the blob suffers when is moved downstream the relativistic jet. 
With this correction, we find
a time delay $\Delta t_{ap} \sim 6.4$ days between the neutrino and the VHE  emission
which  is 
consistent with 
the observations 
\citep[][]{neutrinosmm}.


Despite the simplicity of our model, the results suggest that the observed MM emission of TXS 0506+056 could be explained by the sequence of SED profiles obtained from magnetic reconnection power dissipation and acceleration occurring approximately within the distance range of $\sim$ 2 to 4 pc  along the jet, away from the central BH. 

Future refinement of this model should  include a more realistic background jet provided by numerical relativistic MHD simulations  \citep[e.g.][]{Medina-Torrejón_2021, Medina-Torrejón_2023, medina-torrejon_dalpino2024}, as well as the inclusion of radiative transfer and cosmic ray cascading \citep[as in e.g.][]{rodriguezramires_etal_18, Hussain2023, Hussain2024} to produce  more detailed  SED description. 
With this refinement, it will also be  possible to try to explore the production of the 2014–2015 13$\pm$5 neutrino events,  
likely occurring in the closer to the BH, denser regions of the blazar jet, where high energy photons may be atenuated, possibly explaining  the apparent lack of causal connection of neutrino and EM fluxes during this period.



\section*{Acknowledgements}
 EMdGDP and JCRR  acknowledge support from the Sao Paulo State Funding Agency FAPESP (grants 2013/10559-5, 2017/12188-5, and  2021/02120-0). JCRR also acknowledges support from Rio de Janeiro State Funding Agency FAPERJ (grant E-26/205.635/2022) and EMdGDP the support from CNPq (grant 308643/2017-8). MVdV is supported by FAPESP Grants 2019/05757-9 and 2020/08729-3.  JCRR would like to thank Pankaj Kushwaha for useful
discussions at the first steps of this work.
JCRR made use of Sci-Mind servers machines developed by the CBPF AI LAB team and would like to thank C. R. Bom, P. Russano and M. Portes de Albuquerque for all the support in infrastructure matters.

\section*{Data Availability}
The simulated data generated during this study are available upon
request to the authors.



\bibliographystyle{mnras}
\bibliography{refs} 

\begin{thebibliography}{}
\makeatletter
\relax
\def\mn@urlcharsother{\let\do\@makeother \do\$\do\&\do\#\do\^\do\_\do\%\do\~}
\def\mn@doi{\begingroup\mn@urlcharsother \@ifnextchar [ {\mn@doi@} {\mn@doi@[]}}
\def\mn@doi@[#1]#2{\def\@tempa{#1}\ifx\@tempa\@empty \href {http://dx.doi.org/#2} {doi:#2}\else \href {http://dx.doi.org/#2} {#1}\fi \endgroup}
\def\mn@eprint#1#2{\mn@eprint@#1:#2::\@nil}
\def\mn@eprint@arXiv#1{\href {http://arxiv.org/abs/#1} {{\tt arXiv:#1}}}
\def\mn@eprint@dblp#1{\href {http://dblp.uni-trier.de/rec/bibtex/#1.xml} {dblp:#1}}
\def\mn@eprint@#1:#2:#3:#4\@nil{\def\@tempa {#1}\def\@tempb {#2}\def\@tempc {#3}\ifx \@tempc \@empty \let \@tempc \@tempb \let \@tempb \@tempa \fi \ifx \@tempb \@empty \def\@tempb {arXiv}\fi \@ifundefined {mn@eprint@\@tempb}{\@tempb:\@tempc}{\expandafter \expandafter \csname mn@eprint@\@tempb\endcsname \expandafter{\@tempc}}}

\bibitem[\protect\citeauthoryear{{Abbasi} et~al.,}{{Abbasi} et~al.}{2022}]{Abbasi2022}
{Abbasi} R.,  et~al., 2022, \mn@doi [\apj] {10.3847/1538-4357/ac8de4}, \href {https://ui.adsabs.harvard.edu/abs/2022ApJ...938...38A} {938, 38}

\bibitem[\protect\citeauthoryear{{Ackermann} et~al.,}{{Ackermann} et~al.}{2016}]{ackermann_etal_2016}
{Ackermann} M.,  et~al., 2016, \mn@doi [\apjl] {10.3847/2041-8205/824/2/L20}, \href {https://ui.adsabs.harvard.edu/abs/2016ApJ...824L..20A} {824, L20}

\bibitem[\protect\citeauthoryear{{Agaronyan}, {Atoyan}  \& {Nagapetyan}}{{Agaronyan} et~al.}{1983}]{Agaronyan_1983}
{Agaronyan} F.~A.,  {Atoyan} A.~M.,   {Nagapetyan} A.~M.,  1983, \mn@doi [Astrophysics] {10.1007/BF01005624}, \href {https://ui.adsabs.harvard.edu/abs/1983Ap.....19..187A} {19, 187}

\bibitem[\protect\citeauthoryear{{Aharonian} et~al.,}{{Aharonian} et~al.}{2007}]{aharonian_etal_07}
{Aharonian} F.,  et~al., 2007, \mn@doi [\apjl] {10.1086/520635}, \href {https://ui.adsabs.harvard.edu/abs/2007ApJ...664L..71A} {664, L71}

\bibitem[\protect\citeauthoryear{{Aharonian}, {Kelner}  \& {Prosekin}}{{Aharonian} et~al.}{2010}]{Aharonian_2010}
{Aharonian} F.~A.,  {Kelner} S.~R.,   {Prosekin} A.~Y.,  2010, \mn@doi [\prd] {10.1103/PhysRevD.82.043002}, \href {https://ui.adsabs.harvard.edu/abs/2010PhRvD..82d3002A} {82, 043002}

\bibitem[\protect\citeauthoryear{{Akharonian}, {Kririllov-Ugriumov}  \& {Vardanian}}{{Akharonian} et~al.}{1985}]{Aharonian_1985}
{Akharonian} F.~A.,  {Kririllov-Ugriumov} V.~G.,   {Vardanian} V.~V.,  1985, \mn@doi [\apss] {10.1007/BF00653800}, \href {https://ui.adsabs.harvard.edu/abs/1985Ap&SS.115..201A} {115, 201}

\bibitem[\protect\citeauthoryear{{Ansoldi} et~al.,}{{Ansoldi} et~al.}{2018}]{Ansoldi_2018}
{Ansoldi} S.,  et~al., 2018, \mn@doi [\apjl] {10.3847/2041-8213/aad083}, \href {https://ui.adsabs.harvard.edu/abs/2018ApJ...863L..10A} {863, L10}

\bibitem[\protect\citeauthoryear{{Atoyan} \& {Dermer}}{{Atoyan} \& {Dermer}}{2003}]{Atoyan2003}
{Atoyan} A.~M.,  {Dermer} C.~D.,  2003, \mn@doi [\apj] {10.1086/346261}, \href {https://ui.adsabs.harvard.edu/abs/2003ApJ...586...79A} {586, 79}

\bibitem[\protect\citeauthoryear{{Banik} \& {Bhadra}}{{Banik} \& {Bhadra}}{2019}]{banik2019}
{Banik} P.,  {Bhadra} A.,  2019, \mn@doi [\prd] {10.1103/PhysRevD.99.103006}, \href {https://ui.adsabs.harvard.edu/abs/2019PhRvD..99j3006B} {99, 103006}

\bibitem[\protect\citeauthoryear{{Bellenghi}, {Padovani}, {Resconi}  \& {Giommi}}{{Bellenghi} et~al.}{2023}]{Bellenghi2023}
{Bellenghi} C.,  {Padovani} P.,  {Resconi} E.,   {Giommi} P.,  2023, \mn@doi [\apjl] {10.3847/2041-8213/acf711}, \href {https://ui.adsabs.harvard.edu/abs/2023ApJ...955L..32B} {955, L32}

\bibitem[\protect\citeauthoryear{{Beresnyak} \& {Li}}{{Beresnyak} \& {Li}}{2016}]{beresnyak_etal_2016}
{Beresnyak} A.,  {Li} H.,  2016, \mn@doi [\apj] {10.3847/0004-637X/819/2/90}, \href {https://ui.adsabs.harvard.edu/abs/2016ApJ...819...90B} {819, 90}

\bibitem[\protect\citeauthoryear{{Blandford} \& {Znajek}}{{Blandford} \& {Znajek}}{1977}]{1977MNRAS.179..433B}
{Blandford} R.~D.,  {Znajek} R.~L.,  1977, \mn@doi [\mnras] {10.1093/mnras/179.3.433}, \href {https://ui.adsabs.harvard.edu/abs/1977MNRAS.179..433B} {179, 433}

\bibitem[\protect\citeauthoryear{{Blumenthal}}{{Blumenthal}}{1970}]{Blumenthal_1970b}
{Blumenthal} G.~R.,  1970, \mn@doi [\prd] {10.1103/PhysRevD.1.1596}, \href {https://ui.adsabs.harvard.edu/abs/1970PhRvD...1.1596B} {1, 1596}

\bibitem[\protect\citeauthoryear{{Blumenthal} \& {Gould}}{{Blumenthal} \& {Gould}}{1970}]{Blumenthal_1970}
{Blumenthal} G.~R.,  {Gould} R.~J.,  1970, \mn@doi [Reviews of Modern Physics] {10.1103/RevModPhys.42.237}, \href {https://ui.adsabs.harvard.edu/abs/1970RvMP...42..237B} {42, 237}

\bibitem[\protect\citeauthoryear{{Britto}, {Bottacini}, {Lott}, {Razzaque}  \& {Buson}}{{Britto} et~al.}{2016}]{britto_elal_2016}
{Britto} R.~J.,  {Bottacini} E.,  {Lott} B.,  {Razzaque} S.,   {Buson} S.,  2016, \mn@doi [\apj] {10.3847/0004-637X/830/2/162}, \href {https://ui.adsabs.harvard.edu/abs/2016ApJ...830..162B} {830, 162}

\bibitem[\protect\citeauthoryear{{Britzen} et~al.,}{{Britzen} et~al.}{2019}]{britzen_2019}
{Britzen} S.,  et~al., 2019, \mn@doi [\aap] {10.1051/0004-6361/201935422}, \href {https://ui.adsabs.harvard.edu/abs/2019A&A...630A.103B} {630, A103}

\bibitem[\protect\citeauthoryear{{Buson}, {Tramacere}, {Pfeiffer}, {Oswald}, {de Menezes}, {Azzollini}  \& {Ajello}}{{Buson} et~al.}{2022}]{Buson2022}
{Buson} S.,  {Tramacere} A.,  {Pfeiffer} L.,  {Oswald} L.,  {de Menezes} R.,  {Azzollini} A.,   {Ajello} M.,  2022, \mn@doi [\apjl] {10.3847/2041-8213/ac7d5b}, \href {https://ui.adsabs.harvard.edu/abs/2022ApJ...933L..43B} {933, L43}

\bibitem[\protect\citeauthoryear{{Cerruti}}{{Cerruti}}{2020}]{cerruti_2020}
{Cerruti} M.,  2020, in Journal of Physics Conference Series. p. 012094 (\mn@eprint {arXiv} {1912.03666}), \mn@doi{10.1088/1742-6596/1468/1/012094}

\bibitem[\protect\citeauthoryear{{Cerruti}, {Zech}, {Boisson}, {Emery}, {Inoue}  \& {Lenain}}{{Cerruti} et~al.}{2019}]{Cerruti_2019}
{Cerruti} M.,  {Zech} A.,  {Boisson} C.,  {Emery} G.,  {Inoue} S.,   {Lenain} J.~P.,  2019, \mn@doi [\mnras] {10.1093/mnrasl/sly210}, \href {https://ui.adsabs.harvard.edu/abs/2019MNRAS.483L..12C} {483, L12}

\bibitem[\protect\citeauthoryear{{Comisso} \& {Sironi}}{{Comisso} \& {Sironi}}{2018}]{comisso18}
{Comisso} L.,  {Sironi} L.,  2018, \mn@doi [\prl] {10.1103/PhysRevLett.121.255101}, \href {https://ui.adsabs.harvard.edu/abs/2018PhRvL.121y5101C} {121, 255101}

\bibitem[\protect\citeauthoryear{{Davelaar}, {Philippov}, {Bromberg}  \& {Singh}}{{Davelaar} et~al.}{2020}]{davelaar_etal_2020}
{Davelaar} J.,  {Philippov} A.~A.,  {Bromberg} O.,   {Singh} C.~B.,  2020, \mn@doi [\apjl] {10.3847/2041-8213/ab95a2}, \href {https://ui.adsabs.harvard.edu/abs/2020ApJ...896L..31D} {896, L31}

\bibitem[\protect\citeauthoryear{{Dom{\'\i}nguez} et~al.,}{{Dom{\'\i}nguez} et~al.}{2011}]{Dominguez_2011}
{Dom{\'\i}nguez} A.,  et~al., 2011, \mn@doi [\mnras] {10.1111/j.1365-2966.2010.17631.x}, \href {https://ui.adsabs.harvard.edu/abs/2011MNRAS.410.2556D} {410, 2556}

\bibitem[\protect\citeauthoryear{{Donath} et~al.,}{{Donath} et~al.}{2023}]{Donath_2023}
{Donath} A.,  et~al., 2023, \mn@doi [\aap] {10.1051/0004-6361/202346488}, \href {https://ui.adsabs.harvard.edu/abs/2023A&A...678A.157D} {678, A157}

\bibitem[\protect\citeauthoryear{{Gao}, {Fedynitch}, {Winter}  \& {Pohl}}{{Gao} et~al.}{2019}]{Gao_2019}
{Gao} S.,  {Fedynitch} A.,  {Winter} W.,   {Pohl} M.,  2019, \mn@doi [Nature Astronomy] {10.1038/s41550-018-0610-1}, \href {https://ui.adsabs.harvard.edu/abs/2019NatAs...3...88G} {3, 88}

\bibitem[\protect\citeauthoryear{{Garc{\'\i}a-Morillo} \& {Alexakis}}{{Garc{\'\i}a-Morillo} \& {Alexakis}}{2024}]{Morillo2024}
{Garc{\'\i}a-Morillo} M.,  {Alexakis} Alexandros J.,  2024, \mn@doi [arXiv e-prints] {10.48550/arXiv.2406.08951}, \href {https://ui.adsabs.harvard.edu/abs/2024arXiv240608951M} {p. arXiv:2406.08951}

\bibitem[\protect\citeauthoryear{{Gasparyan}, {B{\'e}gu{\'e}}  \& {Sahakyan}}{{Gasparyan} et~al.}{2022}]{gasparian_2022}
{Gasparyan} S.,  {B{\'e}gu{\'e}} D.,   {Sahakyan} N.,  2022, \mn@doi [\mnras] {10.1093/mnras/stab2688}, \href {https://ui.adsabs.harvard.edu/abs/2022MNRAS.509.2102G} {509, 2102}

\bibitem[\protect\citeauthoryear{{Ghisellini}}{{Ghisellini}}{2013}]{Ghisellini_2013}
{Ghisellini} G.,  2013, {Radiative Processes in High Energy Astrophysics}.
 Lecture Notes in Physics Vol. 873, Springer, Berlin, Germany, \mn@doi{10.1007/978-3-319-00612-3}

\bibitem[\protect\citeauthoryear{{Giannios}}{{Giannios}}{2010}]{giannios_10}
{Giannios} D.,  2010, \mn@doi [\mnras] {10.1111/j.1745-3933.2010.00925.x}, \href {https://ui.adsabs.harvard.edu/abs/2010MNRAS.408L..46G} {408, L46}

\bibitem[\protect\citeauthoryear{{Giannios} \& {Uzdensky}}{{Giannios} \& {Uzdensky}}{2019}]{2019MNRAS.484.1378G}
{Giannios} D.,  {Uzdensky} D.~A.,  2019, \mn@doi [\mnras] {10.1093/mnras/stz082}, \href {https://ui.adsabs.harvard.edu/abs/2019MNRAS.484.1378G} {484, 1378}

\bibitem[\protect\citeauthoryear{{Giannios}, {Uzdensky}  \& {Begelman}}{{Giannios} et~al.}{2009}]{giannios_etal_09}
{Giannios} D.,  {Uzdensky} D.~A.,   {Begelman} M.~C.,  2009, \mn@doi [\mnras] {10.1111/j.1745-3933.2009.00635.x}, \href {https://ui.adsabs.harvard.edu/abs/2009MNRAS.395L..29G} {395, L29}

\bibitem[\protect\citeauthoryear{{Ginzburg} \& {Syrovatskii}}{{Ginzburg} \& {Syrovatskii}}{1964}]{Ginzburg_1964}
{Ginzburg} V.~L.,  {Syrovatskii} S.~I.,  1964, {The Origin of Cosmic Rays}

\bibitem[\protect\citeauthoryear{{Guo}, {Li}, {Daughton}, {Kilian}, {Li}, {Liu}, {Yan}  \& {Ma}}{{Guo} et~al.}{2019}]{guo_etal_2019}
{Guo} F.,  {Li} X.,  {Daughton} W.,  {Kilian} P.,  {Li} H.,  {Liu} Y.-H.,  {Yan} W.,   {Ma} D.,  2019, \mn@doi [\apjl] {10.3847/2041-8213/ab2a15}, \href {https://ui.adsabs.harvard.edu/abs/2019ApJ...879L..23G} {879, L23}

\bibitem[\protect\citeauthoryear{{Guo}, {Liu}, {Li}, {Li}, {Daughton}  \& {Kilian}}{{Guo} et~al.}{2020}]{guo_etal_2020}
{Guo} F.,  {Liu} Y.-H.,  {Li} X.,  {Li} H.,  {Daughton} W.,   {Kilian} P.,  2020, \mn@doi [Physics of Plasmas] {10.1063/5.0012094}, \href {https://ui.adsabs.harvard.edu/abs/2020PhPl...27h0501G} {27, 080501}

\bibitem[\protect\citeauthoryear{{Guo} et~al.,}{{Guo} et~al.}{2022}]{guo_etal_2022}
{Guo} F.,  et~al., 2022, \mn@doi [arXiv e-prints] {10.48550/arXiv.2208.03435}, \href {https://ui.adsabs.harvard.edu/abs/2022arXiv220803435G} {p. arXiv:2208.03435}

\bibitem[\protect\citeauthoryear{{Hovatta} \& {Lindfors}}{{Hovatta} \& {Lindfors}}{2019}]{hovatta_lindfors19}
{Hovatta} T.,  {Lindfors} E.,  2019, \mn@doi [\nar] {10.1016/j.newar.2020.101541}, \href {https://ui.adsabs.harvard.edu/abs/2019NewAR..8701541H} {87, 101541}

\bibitem[\protect\citeauthoryear{{Hovatta}, {Valtaoja}, {Tornikoski}  \& {L{\"a}hteenm{\"a}ki}}{{Hovatta} et~al.}{2009}]{Hovatta2009}
{Hovatta} T.,  {Valtaoja} E.,  {Tornikoski} M.,   {L{\"a}hteenm{\"a}ki} A.,  2009, \mn@doi [\aap] {10.1051/0004-6361:200811150}, \href {https://ui.adsabs.harvard.edu/abs/2009A&A...494..527H} {494, 527}

\bibitem[\protect\citeauthoryear{{Hovatta} et~al.,}{{Hovatta} et~al.}{2021}]{Hovatta2021}
{Hovatta} T.,  et~al., 2021, \mn@doi [\aap] {10.1051/0004-6361/202039481}, \href {https://ui.adsabs.harvard.edu/abs/2021A&A...650A..83H} {650, A83}

\bibitem[\protect\citeauthoryear{{Hussain}, {Alves Batista}, {de Gouveia Dal Pino}  \& {Dolag}}{{Hussain} et~al.}{2023}]{Hussain2023}
{Hussain} S.,  {Alves Batista} R.,  {de Gouveia Dal Pino} E.~M.,   {Dolag} K.,  2023, \mn@doi [Nature Communications] {10.1038/s41467-023-38226-w}, \href {https://ui.adsabs.harvard.edu/abs/2023NatCo..14.2486H} {14, 2486}

\bibitem[\protect\citeauthoryear{{Hussain}, {de Gouveia Dal Pino}  \& {Pagliaroli}}{{Hussain} et~al.}{2024}]{Hussain2024}
{Hussain} S.,  {de Gouveia Dal Pino} E.~M.,   {Pagliaroli} G.,  2024, \mn@doi [\apj] {10.3847/1538-4357/ad10a6}, \href {https://ui.adsabs.harvard.edu/abs/2024ApJ...960..124H} {960, 124}

\bibitem[\protect\citeauthoryear{{IceCube Collaboration} et~al.,}{{IceCube Collaboration} et~al.}{2018a}]{neutrinosprior}
{IceCube Collaboration} et~al., 2018a, \mn@doi [Science] {10.1126/science.aat2890}, \href {https://ui.adsabs.harvard.edu/abs/2018Sci...361..147I} {361, 147}

\bibitem[\protect\citeauthoryear{{IceCube Collaboration} et~al.,}{{IceCube Collaboration} et~al.}{2018b}]{neutrinosmm}
{IceCube Collaboration} et~al., 2018b, \mn@doi [Science] {10.1126/science.aat1378}, \href {https://ui.adsabs.harvard.edu/abs/2018Sci...361.1378I} {361, eaat1378}

\bibitem[\protect\citeauthoryear{Kadowaki, de Gouveia Dal~Pino  \& Stone}{Kadowaki et~al.}{2018}]{Kadowaki_2018}
Kadowaki L. H.~S.,  de Gouveia Dal~Pino E.~M.,   Stone J.~M.,  2018, \mn@doi [The Astrophysical Journal] {10.3847/1538-4357/aad4ff}, 864, 52

\bibitem[\protect\citeauthoryear{{Kadowaki}, {de Gouveia Dal Pino}, {Medina-Torrej{\'o}n}, {Mizuno}  \& {Kushwaha}}{{Kadowaki} et~al.}{2021}]{kadowaki_etal_2021}
{Kadowaki} L. H.~S.,  {de Gouveia Dal Pino} E.~M.,  {Medina-Torrej{\'o}n} T.~E.,  {Mizuno} Y.,   {Kushwaha} P.,  2021, \mn@doi [\apj] {10.3847/1538-4357/abee7a}, \href {https://ui.adsabs.harvard.edu/abs/2021ApJ...912..109K} {912, 109}

\bibitem[\protect\citeauthoryear{{Keivani} et~al.,}{{Keivani} et~al.}{2018}]{Keivani_2018}
{Keivani} A.,  et~al., 2018, \mn@doi [\apj] {10.3847/1538-4357/aad59a}, \href {https://ui.adsabs.harvard.edu/abs/2018ApJ...864...84K} {864, 84}

\bibitem[\protect\citeauthoryear{{Kelner} \& {Aharonian}}{{Kelner} \& {Aharonian}}{2008}]{Kelner_2008}
{Kelner} S.~R.,  {Aharonian} F.~A.,  2008, \mn@doi [\prd] {10.1103/PhysRevD.78.034013}, \href {https://ui.adsabs.harvard.edu/abs/2008PhRvD..78c4013K} {78, 034013}

\bibitem[\protect\citeauthoryear{{Khiali} \& {de Gouveia Dal Pino}}{{Khiali} \& {de Gouveia Dal Pino}}{2016}]{Khiali2016}
{Khiali} B.,  {de Gouveia Dal Pino} E.~M.,  2016, \mn@doi [\mnras] {10.1093/mnras/stv2337}, \href {https://ui.adsabs.harvard.edu/abs/2016MNRAS.455..838K} {455, 838}

\bibitem[\protect\citeauthoryear{{Khiali}, {de Gouveia Dal Pino}  \& {del Valle}}{{Khiali} et~al.}{2015}]{khiali_etal_15}
{Khiali} B.,  {de Gouveia Dal Pino} E.~M.,   {del Valle} M.~V.,  2015, \mn@doi [\mnras] {10.1093/mnras/stv248}, \href {http://adsabs.harvard.edu/abs/2015MNRAS.449...34K} {449, 34}

\bibitem[\protect\citeauthoryear{{Kilian}, {Li}, {Guo}  \& {Li}}{{Kilian} et~al.}{2020}]{kilian_etal_2020}
{Kilian} P.,  {Li} X.,  {Guo} F.,   {Li} H.,  2020, \mn@doi [\apj] {10.3847/1538-4357/aba1e9}, \href {https://ui.adsabs.harvard.edu/abs/2020ApJ...899..151K} {899, 151}

\bibitem[\protect\citeauthoryear{{Kowal}, {Lazarian}, {Vishniac}  \& {Otmianowska-Mazur}}{{Kowal} et~al.}{2009}]{kowal_etal_09}
{Kowal} G.,  {Lazarian} A.,  {Vishniac} E.~T.,   {Otmianowska-Mazur} K.,  2009, ApJ, \href {http://adsabs.harvard.edu/abs/2009ApJ...700...63K} {700, 63}

\bibitem[\protect\citeauthoryear{{Kowal}, {de Gouveia Dal Pino}  \& {Lazarian}}{{Kowal} et~al.}{2011}]{kowal_etal_2011}
{Kowal} G.,  {de Gouveia Dal Pino} E.~M.,   {Lazarian} A.,  2011, \mn@doi [\apj] {10.1088/0004-637X/735/2/102}, \href {https://ui.adsabs.harvard.edu/abs/2011ApJ...735..102K} {735, 102}

\bibitem[\protect\citeauthoryear{{Kowal}, {de Gouveia Dal Pino}  \& {Lazarian}}{{Kowal} et~al.}{2012}]{kowal_etal_2012}
{Kowal} G.,  {de Gouveia Dal Pino} E.~M.,   {Lazarian} A.,  2012, \mn@doi [\prl] {10.1103/PhysRevLett.108.241102}, \href {https://ui.adsabs.harvard.edu/abs/2012PhRvL.108x1102K} {108, 241102}

\bibitem[\protect\citeauthoryear{{Kowal}, {Falceta-Gon{\c{c}}alves}, {Lazarian}  \& {Vishniac}}{{Kowal} et~al.}{2020}]{kowal_etal_2020}
{Kowal} G.,  {Falceta-Gon{\c{c}}alves} D.~A.,  {Lazarian} A.,   {Vishniac} E.~T.,  2020, \mn@doi [\apj] {10.3847/1538-4357/ab7a13}, \href {https://ui.adsabs.harvard.edu/abs/2020ApJ...892...50K} {892, 50}

\bibitem[\protect\citeauthoryear{{Lazarian} \& {Vishniac}}{{Lazarian} \& {Vishniac}}{1999}]{lazarian_vishiniac_99}
{Lazarian} A.,  {Vishniac} E.~T.,  1999, \mn@doi [\apj] {10.1086/307233}, \href {http://adsabs.harvard.edu/abs/1999ApJ...517..700L} {517, 700}

\bibitem[\protect\citeauthoryear{{Li}, {Guo}, {Li}  \& {Li}}{{Li} et~al.}{2015}]{li_etal_2015}
{Li} X.,  {Guo} F.,  {Li} H.,   {Li} G.,  2015, \mn@doi [\apjl] {10.1088/2041-8205/811/2/L24}, \href {https://ui.adsabs.harvard.edu/abs/2015ApJ...811L..24L} {811, L24}

\bibitem[\protect\citeauthoryear{{Liodakis} \& {Petropoulou}}{{Liodakis} \& {Petropoulou}}{2020}]{2020ApJ...893L..20L}
{Liodakis} I.,  {Petropoulou} M.,  2020, \mn@doi [\apjl] {10.3847/2041-8213/ab830a}, \href {https://ui.adsabs.harvard.edu/abs/2020ApJ...893L..20L} {893, L20}

\bibitem[\protect\citeauthoryear{{Lister} et~al.,}{{Lister} et~al.}{2019}]{Lister2019}
{Lister} M.~L.,  et~al., 2019, \mn@doi [\apj] {10.3847/1538-4357/ab08ee}, \href {https://ui.adsabs.harvard.edu/abs/2019ApJ...874...43L} {874, 43}

\bibitem[\protect\citeauthoryear{{Liu}, {Hesse}, {Guo}, {Daughton}, {Li}, {Cassak}  \& {Shay}}{{Liu} et~al.}{2017}]{2017PhRvL.118h5101L}
{Liu} Y.-H.,  {Hesse} M.,  {Guo} F.,  {Daughton} W.,  {Li} H.,  {Cassak} P.~A.,   {Shay} M.~A.,  2017, \mn@doi [\prl] {10.1103/PhysRevLett.118.085101}, \href {https://ui.adsabs.harvard.edu/abs/2017PhRvL.118h5101L} {118, 085101}

\bibitem[\protect\citeauthoryear{{Liu}, {Wang}, {Xue}, {Taylor}, {Wang}, {Li}  \& {Yan}}{{Liu} et~al.}{2019}]{Liu_2019}
{Liu} R.-Y.,  {Wang} K.,  {Xue} R.,  {Taylor} A.~M.,  {Wang} X.-Y.,  {Li} Z.,   {Yan} H.,  2019, \mn@doi [\prd] {10.1103/PhysRevD.99.063008}, \href {https://ui.adsabs.harvard.edu/abs/2019PhRvD..99f3008L} {99, 063008}

\bibitem[\protect\citeauthoryear{{Longair}}{{Longair}}{2011}]{Longair_2011}
{Longair} M.~S.,  2011, {High Energy Astrophysics}

\bibitem[\protect\citeauthoryear{{Lyubarsky} \& {Liverts}}{{Lyubarsky} \& {Liverts}}{2008}]{lyubarsky_etal_2008}
{Lyubarsky} Y.,  {Liverts} M.,  2008, \mn@doi [\apj] {10.1086/589640}, \href {https://ui.adsabs.harvard.edu/abs/2008ApJ...682.1436L} {682, 1436}

\bibitem[\protect\citeauthoryear{{Lyutikov}, {Sironi}, {Komissarov}  \& {Porth}}{{Lyutikov} et~al.}{2017}]{lyutikov_etal_2017}
{Lyutikov} M.,  {Sironi} L.,  {Komissarov} S.~S.,   {Porth} O.,  2017, \mn@doi [Journal of Plasma Physics] {10.1017/S002237781700071X}, \href {https://ui.adsabs.harvard.edu/abs/2017JPlPh..83f6302L} {83, 635830602}

\bibitem[\protect\citeauthoryear{{Maximon}}{{Maximon}}{1968}]{Maximon_1968}
{Maximon} L.~C.,  1968, \mn@doi [J. Res. Nat. Bur. Stand.,] {https://doi.org/10.6028/jres.072B.011}, \href {https://doi.org/10.6028/jres.072B.011} {}

\bibitem[\protect\citeauthoryear{{McKinney}, {Tchekhovskoy}  \& {Blandford}}{{McKinney} et~al.}{2012}]{2012MNRAS.423.3083M}
{McKinney} J.~C.,  {Tchekhovskoy} A.,   {Blandford} R.~D.,  2012, \mn@doi [\mnras] {10.1111/j.1365-2966.2012.21074.x}, \href {https://ui.adsabs.harvard.edu/abs/2012MNRAS.423.3083M} {423, 3083}

\bibitem[\protect\citeauthoryear{{Medina-Torrej{\'o}n}, {de Gouveia Dal Pino}, {Kadowaki}, {Kowal}, {Singh}  \& {Mizuno}}{{Medina-Torrej{\'o}n} et~al.}{2021}]{Medina-Torrejón_2021}
{Medina-Torrej{\'o}n} T.~E.,  {de Gouveia Dal Pino} E.~M.,  {Kadowaki} L. H.~S.,  {Kowal} G.,  {Singh} C.~B.,   {Mizuno} Y.,  2021, \mn@doi [\apj] {10.3847/1538-4357/abd6c2}, \href {https://ui.adsabs.harvard.edu/abs/2021ApJ...908..193M} {908, 193}

\bibitem[\protect\citeauthoryear{{Medina-Torrej{\'o}n}, {de Gouveia Dal Pino}  \& {Kowal}}{{Medina-Torrej{\'o}n} et~al.}{2023}]{Medina-Torrejón_2023}
{Medina-Torrej{\'o}n} T.~E.,  {de Gouveia Dal Pino} E.~M.,   {Kowal} G.,  2023, \mn@doi [\apj] {10.3847/1538-4357/acd699}, \href {https://ui.adsabs.harvard.edu/abs/2023ApJ...952..168M} {952, 168}

\bibitem[\protect\citeauthoryear{{M{\"u}cke}, {Engel}, {Rachen}, {Protheroe}  \& {Stanev}}{{M{\"u}cke} et~al.}{2000}]{Mucke_2000}
{M{\"u}cke} A.,  {Engel} R.,  {Rachen} J.~P.,  {Protheroe} R.~J.,   {Stanev} T.,  2000, \mn@doi [Computer Physics Communications] {10.1016/S0010-4655(99)00446-4}, \href {https://ui.adsabs.harvard.edu/abs/2000CoPhC.124..290M} {124, 290}

\bibitem[\protect\citeauthoryear{{Murase}, {Oikonomou}  \& {Petropoulou}}{{Murase} et~al.}{2018}]{murae2018}
{Murase} K.,  {Oikonomou} F.,   {Petropoulou} M.,  2018, \mn@doi [\apj] {10.3847/1538-4357/aada00}, \href {https://ui.adsabs.harvard.edu/abs/2018ApJ...865..124M} {865, 124}

\bibitem[\protect\citeauthoryear{{Nishikawa} et~al.,}{{Nishikawa} et~al.}{2020}]{nishikawa_etal_2020}
{Nishikawa} K.,  et~al., 2020, \mn@doi [\mnras] {10.1093/mnras/staa421}, \href {https://ui.adsabs.harvard.edu/abs/2020MNRAS.493.2652N} {493, 2652}

\bibitem[\protect\citeauthoryear{{Padovani}, {Oikonomou}, {Petropoulou}, {Giommi}  \& {Resconi}}{{Padovani} et~al.}{2019}]{Padovani2019}
{Padovani} P.,  {Oikonomou} F.,  {Petropoulou} M.,  {Giommi} P.,   {Resconi} E.,  2019, \mn@doi [\mnras] {10.1093/mnrasl/slz011}, \href {https://ui.adsabs.harvard.edu/abs/2019MNRAS.484L.104P} {484, L104}

\bibitem[\protect\citeauthoryear{{Paiano}, {Falomo}, {Treves}  \& {Scarpa}}{{Paiano} et~al.}{2018}]{2018ApJ...854L..32P}
{Paiano} S.,  {Falomo} R.,  {Treves} A.,   {Scarpa} R.,  2018, \mn@doi [\apjl] {10.3847/2041-8213/aaad5e}, \href {https://ui.adsabs.harvard.edu/abs/2018ApJ...854L..32P} {854, L32}

\bibitem[\protect\citeauthoryear{{Petropoulou}, {Giannios}  \& {Sironi}}{{Petropoulou} et~al.}{2016}]{petropoulou_etal_2016}
{Petropoulou} M.,  {Giannios} D.,   {Sironi} L.,  2016, \mn@doi [\mnras] {10.1093/mnras/stw1832}, \href {https://ui.adsabs.harvard.edu/abs/2016MNRAS.462.3325P} {462, 3325}

\bibitem[\protect\citeauthoryear{{Petropoulou} et~al.,}{{Petropoulou} et~al.}{2020}]{petropoulou2020}
{Petropoulou} M.,  et~al., 2020, \mn@doi [\apj] {10.3847/1538-4357/ab76d0}, \href {https://ui.adsabs.harvard.edu/abs/2020ApJ...891..115P} {891, 115}

\bibitem[\protect\citeauthoryear{{Plavin}, {Kovalev}, {Kovalev}  \& {Troitsky}}{{Plavin} et~al.}{2020}]{Plavin2020}
{Plavin} A.,  {Kovalev} Y.~Y.,  {Kovalev} Y.~A.,   {Troitsky} S.,  2020, \mn@doi [\apj] {10.3847/1538-4357/ab86bd}, \href {https://ui.adsabs.harvard.edu/abs/2020ApJ...894..101P} {894, 101}

\bibitem[\protect\citeauthoryear{{Plavin}, {Kovalev}, {Kovalev}  \& {Troitsky}}{{Plavin} et~al.}{2021}]{Plavin2021}
{Plavin} A.~V.,  {Kovalev} Y.~Y.,  {Kovalev} Y.~A.,   {Troitsky} S.~V.,  2021, \mn@doi [\apj] {10.3847/1538-4357/abceb8}, \href {https://ui.adsabs.harvard.edu/abs/2021ApJ...908..157P} {908, 157}

\bibitem[\protect\citeauthoryear{{Rees}}{{Rees}}{1966}]{Rees1966}
{Rees} M.~J.,  1966, \mn@doi [\nat] {10.1038/211468a0}, \href {https://ui.adsabs.harvard.edu/abs/1966Natur.211..468R} {211, 468}

\bibitem[\protect\citeauthoryear{{Reimer}, {B{\"o}ttcher}  \& {Buson}}{{Reimer} et~al.}{2019}]{reimer2019}
{Reimer} A.,  {B{\"o}ttcher} M.,   {Buson} S.,  2019, \mn@doi [\apj] {10.3847/1538-4357/ab2bff}, \href {https://ui.adsabs.harvard.edu/abs/2019ApJ...881...46R} {881, 46}

\bibitem[\protect\citeauthoryear{{Ripperda}, {Porth}, {Xia}  \& {Keppens}}{{Ripperda} et~al.}{2017}]{Ripperda2017}
{Ripperda} B.,  {Porth} O.,  {Xia} C.,   {Keppens} R.,  2017, \mn@doi [\mnras] {10.1093/mnras/stx1875}, \href {https://ui.adsabs.harvard.edu/abs/2017MNRAS.471.3465R} {471, 3465}

\bibitem[\protect\citeauthoryear{{Rodrigues}, {Gao}, {Fedynitch}, {Palladino}  \& {Winter}}{{Rodrigues} et~al.}{2019}]{rodrigues2019}
{Rodrigues} X.,  {Gao} S.,  {Fedynitch} A.,  {Palladino} A.,   {Winter} W.,  2019, \mn@doi [\apjl] {10.3847/2041-8213/ab1267}, \href {https://ui.adsabs.harvard.edu/abs/2019ApJ...874L..29R} {874, L29}

\bibitem[\protect\citeauthoryear{{Rodriguez-Ramirez}, {de Gouveia Dal Pino}  \& {Alves Batista}}{{Rodriguez-Ramirez} et~al.}{2019}]{rodriguezramires_etal_18}
{Rodriguez-Ramirez} J.~C.,  {de Gouveia Dal Pino} E.~M.,   {Alves Batista} R.,  2019, \mn@doi [\apj] {10.3847/1538-4357/ab212e}, \href {https://ui.adsabs.harvard.edu/abs/2019ApJ...879....6R} {879, 6}

\bibitem[\protect\citeauthoryear{{Romero}, {Vieyro}  \& {Vila}}{{Romero} et~al.}{2010}]{Romero2010}
{Romero} G.~E.,  {Vieyro} F.~L.,   {Vila} G.~S.,  2010, \mn@doi [\aap] {10.1051/0004-6361/200913663}, \href {https://ui.adsabs.harvard.edu/abs/2010A&A...519A.109R} {519, A109}

\bibitem[\protect\citeauthoryear{{Sahakyan}}{{Sahakyan}}{2018}]{Sahakyan_2018}
{Sahakyan} N.,  2018, \mn@doi [\apj] {10.3847/1538-4357/aadade}, \href {https://ui.adsabs.harvard.edu/abs/2018ApJ...866..109S} {866, 109}

\bibitem[\protect\citeauthoryear{{Singh}, {Mizuno}  \& {de Gouveia Dal Pino}}{{Singh} et~al.}{2016}]{singh_etal_16}
{Singh} C.~B.,  {Mizuno} Y.,   {de Gouveia Dal Pino} E.~M.,  2016, \mn@doi [\apj] {10.3847/0004-637X/824/1/48}, \href {http://adsabs.harvard.edu/abs/2016ApJ...824...48S} {824, 48}

\bibitem[\protect\citeauthoryear{{Sironi} \& {Spitkovsky}}{{Sironi} \& {Spitkovsky}}{2014}]{sironi_spitkovsky_2014}
{Sironi} L.,  {Spitkovsky} A.,  2014, \mn@doi [\apjl] {10.1088/2041-8205/783/1/L21}, \href {https://ui.adsabs.harvard.edu/abs/2014ApJ...783L..21S} {783, L21}

\bibitem[\protect\citeauthoryear{{Takamoto}, {Inoue}  \& {Lazarian}}{{Takamoto} et~al.}{2015}]{takamoto_etal_15}
{Takamoto} M.,  {Inoue} T.,   {Lazarian} A.,  2015, \mn@doi [\apj] {10.1088/0004-637X/815/1/16}, \href {https://ui.adsabs.harvard.edu/abs/2015ApJ...815...16T} {815, 16}

\bibitem[\protect\citeauthoryear{{Vicentin}, {Kowal}, {de Gouveia Dal Pino}  \& {Lazarian}}{{Vicentin} et~al.}{2024}]{Vicentin2024}
{Vicentin} G.~H.,  {Kowal} G.,  {de Gouveia Dal Pino} E.~M.,   {Lazarian} A.,  2024, \mn@doi [arXiv e-prints] {10.48550/arXiv.2405.15909}, \href {https://ui.adsabs.harvard.edu/abs/2024arXiv240515909V} {p. arXiv:2405.15909}

\bibitem[\protect\citeauthoryear{{Werner}, {Philippov}  \& {Uzdensky}}{{Werner} et~al.}{2019}]{werner_etal_2019}
{Werner} G.~R.,  {Philippov} A.~A.,   {Uzdensky} D.~A.,  2019, \mn@doi [\mnras] {10.1093/mnrasl/sly157}, \href {https://ui.adsabs.harvard.edu/abs/2019MNRAS.482L..60W} {482, L60}

\bibitem[\protect\citeauthoryear{{Xu} \& {Lazarian}}{{Xu} \& {Lazarian}}{2023}]{xu_lazarian_2023}
{Xu} S.,  {Lazarian} A.,  2023, \mn@doi [\apj] {10.3847/1538-4357/aca32c}, \href {https://ui.adsabs.harvard.edu/abs/2023ApJ...942...21X} {942, 21}

\bibitem[\protect\citeauthoryear{{Xue}, {Liu}, {Petropoulou}, {Oikonomou}, {Wang}, {Wang}  \& {Wang}}{{Xue} et~al.}{2019}]{Xue_2019}
{Xue} R.,  {Liu} R.-Y.,  {Petropoulou} M.,  {Oikonomou} F.,  {Wang} Z.-R.,  {Wang} K.,   {Wang} X.-Y.,  2019, \mn@doi [\apj] {10.3847/1538-4357/ab4b44}, \href {https://ui.adsabs.harvard.edu/abs/2019ApJ...886...23X} {886, 23}

\bibitem[\protect\citeauthoryear{{Zdziarski} \& {Bottcher}}{{Zdziarski} \& {Bottcher}}{2015}]{2015MNRAS.450L..21Z}
{Zdziarski} A.~A.,  {Bottcher} M.,  2015, \mn@doi [\mnras] {10.1093/mnrasl/slv039}, \href {https://ui.adsabs.harvard.edu/abs/2015MNRAS.450L..21Z} {450, L21}

\bibitem[\protect\citeauthoryear{{Zhang}, {Li}, {Guo}  \& {Giannios}}{{Zhang} et~al.}{2018}]{zhang_etal_2018}
{Zhang} H.,  {Li} X.,  {Guo} F.,   {Giannios} D.,  2018, \mn@doi [\apjl] {10.3847/2041-8213/aad54f}, \href {https://ui.adsabs.harvard.edu/abs/2018ApJ...862L..25Z} {862, L25}

\bibitem[\protect\citeauthoryear{{Zhang}, {Petropoulou}, {Murase}  \& {Oikonomou}}{{Zhang} et~al.}{2020}]{zhang_murase2020}
{Zhang} B.~T.,  {Petropoulou} M.,  {Murase} K.,   {Oikonomou} F.,  2020, \mn@doi [\apj] {10.3847/1538-4357/ab659a}, \href {https://ui.adsabs.harvard.edu/abs/2020ApJ...889..118Z} {889, 118}

\bibitem[\protect\citeauthoryear{{Zhang}, {Xu}, {Lazarian}  \& {Kowal}}{{Zhang} et~al.}{2023a}]{Zhang_Xu2023}
{Zhang} J.-F.,  {Xu} S.,  {Lazarian} A.,   {Kowal} G.,  2023a, \mn@doi [Journal of High Energy Astrophysics] {10.1016/j.jheap.2023.08.001}, \href {https://ui.adsabs.harvard.edu/abs/2023JHEAp..40....1Z} {40, 1}

\bibitem[\protect\citeauthoryear{{Zhang}, {Sironi}, {Giannios}  \& {Petropoulou}}{{Zhang} et~al.}{2023b}]{Zhang2023}
{Zhang} H.,  {Sironi} L.,  {Giannios} D.,   {Petropoulou} M.,  2023b, \mn@doi [\apjl] {10.3847/2041-8213/acfe7c}, \href {https://ui.adsabs.harvard.edu/abs/2023ApJ...956L..36Z} {956, L36}

\bibitem[\protect\citeauthoryear{{de Gouveia Dal Pino} \& {Kowal}}{{de Gouveia Dal Pino} \& {Kowal}}{2015}]{dalpino_kowal_15}
{de Gouveia Dal Pino} E.~M.,  {Kowal} G.,  2015, {Particle Acceleration by Magnetic Reconnection}.
Springer Berlin Heidelberg, p.~373, \mn@doi{10.1007/978-3-662-44625-6_13}

\bibitem[\protect\citeauthoryear{{de Gouveia Dal Pino} \& {Lazarian}}{{de Gouveia Dal Pino} \& {Lazarian}}{2005}]{dalpino_lazarian_2005}
{de Gouveia Dal Pino} E.~M.,  {Lazarian} A.,  2005, \mn@doi [\aap] {10.1051/0004-6361:20042590}, \href {https://ui.adsabs.harvard.edu/abs/2005A&A...441..845D} {441, 845}

\bibitem[\protect\citeauthoryear{{de Gouveia Dal Pino} \& {Medina-Torrejon}}{{de Gouveia Dal Pino} \& {Medina-Torrejon}}{2024}]{medina-torrejon_dalpino2024}
{de Gouveia Dal Pino} E.~M.,  {Medina-Torrejon} T.~E.,  2024, \mn@doi [arXiv e-prints] {10.48550/arXiv.2410.13071}, \href {https://ui.adsabs.harvard.edu/abs/2024arXiv241013071D} {p. arXiv:2410.13071}

\bibitem[\protect\citeauthoryear{{de Gouveia Dal Pino}, {Kowal}, {Kadowaki}, {Medina-Torrej{\'o}n}, {Mizuno}  \& {Singh}}{{de Gouveia Dal Pino} et~al.}{2020}]{dalpino_etal_2020}
{de Gouveia Dal Pino} E.~M.,  {Kowal} G.,  {Kadowaki} L.,  {Medina-Torrej{\'o}n} T.~E.,  {Mizuno} Y.,   {Singh} C.,  2020, in {Asada} K.,  {de Gouveia Dal Pino} E.,  {Giroletti} M.,  {Nagai} H.,   {Nemmen} R.,  eds,  IAU Symposium Vol. 342, IAU Symposium. pp 13--18, \mn@doi{10.1017/S1743921318007688}

\bibitem[\protect\citeauthoryear{{de Gouveia dal Pino} \& {Opher}}{{de Gouveia dal Pino} \& {Opher}}{1991}]{dalpino1991}
{de Gouveia dal Pino} E.~M.,  {Opher} R.,  1991, \aap, \href {https://ui.adsabs.harvard.edu/abs/1991A&A...242..319D} {242, 319}

\bibitem[\protect\citeauthoryear{{del Valle}, {de Gouveia Dal Pino}  \& {Kowal}}{{del Valle} et~al.}{2016}]{delvalle_etal_16}
{del Valle} M.~V.,  {de Gouveia Dal Pino} E.~M.,   {Kowal} G.,  2016, \mn@doi [\mnras] {10.1093/mnras/stw2276}, \href {http://adsabs.harvard.edu/abs/2016MNRAS.463.4331D} {463, 4331}

\makeatother
\end{thebibliography}


\newpage

\section{appendix A: 
Time scales of particle 
energy losses}
\label{sec:Appendix A}

The cooling time scale of a particle $i$(=proton, electron, positron) with energy $E_i$ is taken as the inverse of the rate of the cooling process. 

In the present emission model, the cooling rate (s$^{-1}$) of a charged particle $i$  due to synchrotron radiation  is calculated as:
\begin{equation}
t^{-1}_\mathrm{syn}(E_i) = \frac{ P_\mathrm{syn,tot}}{E_i},
\end{equation}
with 
\begin{equation}
P_\mathrm{syn,tot} = \frac{4}{3}  \left(\frac{m_\mathrm{e}}{m_i}\right)^2  \sigma_\mathrm{T}  c U_\mathrm{B}
\left(\frac{E_i}{m_i c^2}\right)^2,
\end{equation}
being the radiated power of the particle, and $U_\mathrm{B}=B^2/(8\pi)$ magnetic energy density of the of the emission region.

We also consider the inverse compton (IC) scattering for electrons and positrons, for which the rate is calculated as
\begin{equation}
t^{-1}_\mathrm{IC} = 
\frac{1}{E_\mathrm{e}}
\int_{\epsilon_\mathrm{s,m}}^{\epsilon_\mathrm{s,M}} d\epsilon_\mathrm{s}
\int_{\epsilon_\mathrm{s}}^{\frac{g E_\mathrm{e}}{1+g}} dE_\gamma,
\frac{2\pi r_0^2 m_\mathrm{e}^2 c^5}{E_\mathrm{e}^2}
\frac{n_\mathrm{ph}^\mathrm{syn}{}}{\epsilon_\mathrm{s}} F(\epsilon_\mathrm{c},\epsilon_\mathrm{syn},E_\mathrm{e})
\end{equation}
with $g =4 \epsilon_\mathrm{s}E_\mathrm{e}/(m_\mathrm{e} c^2)^2$, and
\begin{align}
\nonumber
& F(\epsilon_\mathrm{c},\epsilon_\mathrm{syn},E_\mathrm{e}) = 2q \ln (q)  + (1+2q)(1-q) + \frac{(1-q)b^2}{2(1 + b)},\\
& q=\frac{\epsilon_c (m_\mathrm{e} c^2)^2}{ 4 \epsilon_s E'_e (E'_e - \epsilon_c) },\,\,
b=\frac{\epsilon_\mathrm{c}}{( E'_\mathrm{e} - \epsilon_\mathrm{c})}.
\label{FIC}
\end{align}

 
Protons also cool due to their interactions with the target soft radiation, producing positron pairs, thorough the Bethe-Heitler (B-H) process, and producing pions through the photo-pion channel.
 Both processes occur when the target photons surpass the threshold energies
of $\epsilon_\mathrm{0,BH}=2m_e c^2\approx1.02$ MeV and $\epsilon_{0,\pi}=145$ MeV, respectively, 
in the rest frame of the relativistic proton. 
Here we calculate the cooling rate due to 
these photo-hadronic processes as
\citep[e.g.,][]{Atoyan2003, Romero2010, khiali_etal_15}:
\begin{equation}
    t_{p\gamma}^{-1}(E_p) = 
    \frac{c}{2 \gamma_\mathrm{p}^2}
    \int_{\frac{\epsilon_0}{2\gamma_p}}^{\infty} d\epsilon \frac{n^\mathrm{syn}_\mathrm{ph}(\epsilon)}{\epsilon^2}
    \int_{\epsilon_0}^{2\epsilon\gamma_p} d\epsilon' \epsilon' 
    K_{p\gamma}\sigma_{p\gamma},
\label{rate_pg}
\end{equation}
where $\sigma_{p\gamma}$, and $K_{p\gamma}=$ are the cross section and the inelasticity of the process.
For the case of the B-H process we employ equation (\ref{rate_pg}) using 
$\epsilon_0 =\epsilon_\mathrm{0,BH}$, 
$K_\mathrm{p\gamma}=2m_\mathrm{e}/m_\mathrm{p}$, and
$\sigma_\mathrm{p\gamma}$ is taken from \cite{Maximon_1968}.
For the photo-pion interaction we calculate the cooling rate using equation (\ref{rate_pg})
with
$\epsilon_0 =\epsilon_\mathrm{0,\pi}$, and 
$K_\mathrm{p\gamma}$, and
$\sigma_\mathrm{p\gamma}$ are obtained following the approach of \cite{Atoyan2003}.

\section{appendix B:
Particle emission fluxes}
\label{sec:Appendix B}

We derive the MM emission that results from electrons and protons accelerated within the reconnection region in the blazar jet, neglecting the influence of any external radiation field (such as the one produced by the AGN disk, and/or a broad line region).
Thus, we consider the Synchrotron radiation of accelerated electrons  as the only source of target photons within the emission volume leading to SSC scattering, photo-pion production, Bethe-Heitler pair production, and pair production by photon-photon anihilation.
In what follows, we describe the particle and radiation processes that we account for, with physical quantities defined in the co-moving frame of the emission region, unless we specify quantities in a different frame.

We consider the radiation fields of low and high energy photons as isotropic an uniform  within a spherical region of radius $r_\mathrm{b}$
and approximate the energy distributions of photon field densities (in erg$^{-1}$ cm$^{-3}$) as:
\begin{equation}
n_\mathrm{ph}(\epsilon) = \frac{4\pi}{c\epsilon} I_\epsilon.
\label{nph_gral}
\end{equation}
In this equation,  $I_\epsilon$ is the specific intensity of the radiation field defined as $dE=I_\epsilon dt d\epsilon dA d\Omega$ (s$^{-1}$ cm$^{-2}$ sr$^{-1}$) at a given photon energy 
$\epsilon$.
The simplest solution for $I_\epsilon$ in an uniform medium with emission and absorption can be written as:
\begin{equation}
I_\epsilon = j_\epsilon r_\mathrm{b}
\frac{1-e^{-\tau} }{\tau},
\label{I_eps}
\end{equation}
being $j_\epsilon$ the emission coefficient,  
$\tau\approx r_\mathrm{b}\alpha_\epsilon$ 
the optical depth, and 
$\alpha_\epsilon$ the corresponding absorption coefficient.

\subsection{Leptonic emission}
\label{lepton}

Synchrotron photons emitted by accelerated electrons are assumed here as the source of target photons within the emission region as well as the source of the low energy bump in the observed blazar SED.
We approximate the photon field density of this Synchroton emission neglecting the effect of Synchrotron self-absorption:
\begin{equation}
n^\mathrm{syn}_\mathrm{ph}(\epsilon) = \frac{4\pi  j^\mathrm{syn}_{\epsilon} r_\mathrm{b}}{c\epsilon}, 
\label{nph}
\end{equation}
where $j^\mathrm{syn}_{\epsilon}$ is the Synchrotron emission coefficient (s$^{-1}$ cm$^{-3}$ sr$^{-1}$ Hz$^{-1}$) at the photon energy $\epsilon$. 
Given the energy distribution of primary electrons $N'_\mathrm{e}$ (modelled by eq. \ref{Ne_bpl}), we obtain
the Synchrotron emission coefficient as
\begin{equation}
4\pi j^\mathrm{syn}_{\epsilon} = \int_{E'_\mathrm{e,0}}^{\infty}
dE''_\mathrm{e} N_\mathrm{e}(E''_\mathrm{e}) \langle P_\mathrm{syn}(E''_\mathrm{e})\rangle_\mathrm{\alpha},
\label{j_eps_syn}
\end{equation}
where $\langle P_\mathrm{syn} \rangle_\alpha$ is the angular averaged power of radiation emitted by a single electron of energy $E'_\mathrm{e}$ in a random magnetic field, which we calculate using the the approach of \cite{Aharonian_2010} (their Appendix D).
The differential luminosity of Synchrotron photons (luminosity per photon energy) that leaves the emission region is then calculated as:
\begin{equation}
L^\mathrm{syn}_{\epsilon} = V_\mathrm{b} 4\pi j^\mathrm{syn}_{\epsilon},
\label{Lsyn}
\end{equation}
where we aproximate the reconnection volume as $V_\mathrm{b}=4\pi r_\mathrm{b}^3/3$.

We account for the SSC emission of the accelerated electrons as:
\begin{equation}
L^\mathrm{ssc}_{\epsilon}= 
V_\mathrm{b} 4\pi j^\mathrm{IC}_{\epsilon}
\left[1- \exp\{- \tau_{\gamma \gamma}(\epsilon)\} \right]
/\tau_{\gamma \gamma} (\epsilon) 
\label{L_e_C}
\end{equation}
where $j^\mathrm{IC}_{\epsilon}$ is the IC emission coefficient,
and 
\begin{equation}
\tau_{\gamma\gamma} (\epsilon)= r_\mathrm{b} \int_\mathrm{\epsilon_{s,min}}^{\epsilon_\mathrm{s,max}} 
d\epsilon_\mathrm{s}
n^\mathrm{syn}_\mathrm{ph}(\epsilon_\mathrm{s})\sigma_{\gamma \gamma}(\epsilon_\mathrm{s},\epsilon),
\label{tau_gg}
\end{equation}
is the optical depth due to gamma-ray annihilation by pair production within the emission volume.
Given the energy distribution of primary electrons $N'_\mathrm{e}$ (eq. \ref{Ne_bpl}), and the energy density distribution of the target soft photons $n^\mathrm{syn}_\mathrm{ph}(\epsilon_\mathrm{s})$ (eq. \ref{nph}), we calculate the 
IC scattering emission coefficient  
following the approach of \cite{Blumenthal_1970} which allows for Thompson and Klein-Nishina regimes:
\begin{equation}
4\pi j^\mathrm{IC}_{\epsilon}=
\epsilon
\int_{E'_\mathrm{e,0}}^{\infty} dE_\mathrm{e} N'_\mathrm{e}(E'_\mathrm{e})
\int_\mathrm{\epsilon_\mathrm{s,0}}^{\epsilon_\mathrm{s,max}} d\epsilon_\mathrm{syn} n^\mathrm{syn}_\mathrm{ph}(\epsilon_\mathrm{s})
K_\mathrm{IC}(\epsilon,\epsilon_\mathrm{s},E'_\mathrm{e}),
\end{equation}
with
\begin{equation}
K_\mathrm{IC}(\epsilon,\epsilon_\mathrm{s},E_\mathrm{e}) =
\frac{3\sigma_\mathrm{T} (m_\mathrm{e} c^2)^2}{4\epsilon_\mathrm{s} E_\mathrm{e}^2}
F(\epsilon,\epsilon_\mathrm{s},E_\mathrm{e}),
\end{equation}
and $F(\epsilon_\mathrm{c},\epsilon_\mathrm{s},E_\mathrm{e})$ is given by equation (\ref{FIC}).
To obtain $\tau_{\gamma\gamma}$ we use the cross section for photon-photon annihilation $\sigma_{\gamma\gamma}$ (see, e.g., \citealt{Aharonian_1985}) in the head-on collission limit: 
\begin{equation}
\sigma_{\gamma \gamma} = 
\frac{1}{2}\pi r_\mathrm{e}^2
\left(
1-\beta_0^2
\right)
\left[
\left(3-\beta_0^4\right) \ln\left\{\frac{1+\beta_0}{1-\beta_0}\right\}
+2\beta_0\left(\beta_0^2-2\right)
\right],
\end{equation}
with
\begin{equation}
\beta_0 = \sqrt{1-\frac{m_\mathrm{e}^2c^4}{\epsilon_\mathrm{s}\epsilon}},
\end{equation}
where $\epsilon_\mathrm{s}$ and $\epsilon$ are the energies of the low and high energy photons, respectively.

\subsection{Hadronic dissipation}
In the reconnection region analysed here, we consider proton Synchrotron radiation, photon-pion production, and Bethe-Heitler pair production as the dominant hadronic processes that contribute to observable emission from the blazar flare.

We obtain the differential luminosity due to proton Synchrotron emission similarly as in the case of primary electrons (see eq. \ref{Lsyn}) but employing 
the energy distribution of protons derived in eq. (\ref{Np}) as well as the proton mass $m_\mathrm{p}$ when evaluating the emission coefficient
 $j^\mathrm{syn}_\epsilon$ (detailed in section \ref{lepton}).

For the photo-pion cooling channel,
we account for the injection of particles and photons within the emission region employing the formalism of \cite{Kelner_2008} which provides analytical approximations from Monte Carlo simulations performed with the \texttt{SOPHIA} code \citep{Mucke_2000}.
Given the population of relativistic protons (eq. \ref{Np}),
and the distribution of soft photons (eq. \ref{nph}),  we account for the injection rate $q_\ell$
(s$^{-1}$ erg$^{-3}$ cm$^{-3}$)
of the gamma-rays ($\gamma$), neutrinos($\nu$), anti-neutrinos ($\bar{\nu}$), electrons (e$^{-}$), and positrons (e$^{+}$)
evaluating the expressions:
\begin{align}
\nonumber
q_{\ell}(E_\ell) = & 
\int^{\infty}_{\eta_0} d\eta
H(\eta, E_\ell),\\
H(\eta, E_\ell) = &
\frac{m_\mathrm{p}^2 c^4}{4}
\int^{\infty}_{E_\ell}
\frac{dE_\mathrm{p}}{E^2_\mathrm{p}} N_\mathrm{p} (E_\mathrm{p})
n_\mathrm{ph}\left(\frac{\eta m^2_\mathrm{p}c^4}{4E_\mathrm{p}}\right)
\Phi_\ell\left(\eta, \frac{E_\ell}{E_\mathrm{p}}\right),
\label{q_ell}
\end{align}
where
$\eta_0 = 2\frac{m_\pi}{m_\mathrm{p}} + \frac{m_\pi^2}{m_\mathrm{p}^2}\approx 0.313 $,
the sub-index $\ell$ labels the products 
$\ell= \gamma, \nu, \bar{\nu}, e^+, e^-$, and $\Phi_\ell$ is the energy distribution of the type of product $\ell$ given by equations 27 and 31, and tables I, II and III in \cite{Kelner_2008}.

Once produced, neutrinos are virtually unimpeded until their potential detection at Earth. 
We then calculate the differential luminosity of, e.g., 
muon neutrinos leaving the emission region as:
\begin{equation}
L_{E_\nu} = V_\mathrm{b} E_\nu \left(
q_{\nu_\mu} + q_{\bar{\nu}_{\mu}}
\right),
\label{Lnus}
\end{equation}
where $q_{\nu_\mu}$ and $q_{\bar{\nu}_\mu}$ are obtained from eq. \ref{q_ell}.
The electrons, positrons and gamma-rays injected from the decay of pions in the photo-hadronic channel (see eq. \ref{q_ell}), manifest as observable radiation via electromagnetic cascades within the emission region. We account for such reprocessed radiation as detailed in the next subsection.

When the energy of relativistic protons 
$E_\mathrm{p} = \gamma_\mathrm{p} m_\mathrm{p} c^2$ and of target soft photons $\epsilon$ surpass the threshold 
$\gamma_\mathrm{p}\epsilon>m_\mathrm{e}c^2$,
we expect direct production of electron-prositron pairs
by the \ Bethe-Hitler channel.
We account for the injection rate of these pairs (erg$^{-1}$ s$^{-1}$, cm$^{-3}$) in the emission region as:
\begin{equation}
q^\mathrm{B-H}_\mathrm{e}(E_\mathrm{e})=
\int^{\infty}_{E_\mathrm{p,0}}
dE_\mathrm{p}N_\mathrm{p}(E_\mathrm{p}) \Phi_\mathrm{BH}(E_\mathrm{p},E_\mathrm{e}),
\label{q_BH}
\end{equation}
where
\begin{align}
\nonumber
\Phi_\mathrm{BH}(E_\mathrm{p},E_\mathrm{e}) = \frac{1}{2\gamma^3_\mathrm{p} m_\mathrm{e} c}
&\int^{\infty}_{\frac{(\gamma_\mathrm{p}+\gamma_\mathrm{e})^2}{4\gamma_\mathrm{p}^2\gamma_\mathrm{e}}} d\tilde{\epsilon} \frac{\tilde{n}(\tilde{\epsilon})}{\tilde{\epsilon}^2}\\
\times&\int^{2\gamma_\mathrm{p}\tilde{\epsilon}}_{\frac{(\gamma_\mathrm{p}+\gamma_\mathrm{e})^2}{2\gamma_\mathrm{p}\gamma_\mathrm{e}}} d\omega \omega
\int^{\omega - 1}_{\frac{\gamma^2_\mathrm{p} + \gamma^2_\mathrm{e}}{2\gamma_\mathrm{p}\gamma_\mathrm{e}}} d\tilde{\gamma}_{-}
\frac{W(\omega, \tilde{\gamma}_\mathrm{-},\xi)}{\tilde{p}_{-}},
\label{KA_BH}
\end{align}
is the injection rate of electrons by a proton of energy $E_\mathrm{p}$ as presented in \cite{Kelner_2008}, with  $W$ being the differential cross section of the process in the rest frame of the proton taken from \cite{Blumenthal_1970b}.
In eq. (\ref{KA_BH}),  
$\gamma_\mathrm{e,p}=\frac{E_{\mathrm{e,p}}}{m_\mathrm{e,p} c^2}$ are the Lorentz factor of the  electrons and protons in the emission volume frame,
$\tilde{\epsilon}=\frac{\epsilon}{m_\mathrm{e} c^2} $
and $\tilde{n}=m_\mathrm{e}c^2 n_\mathrm{ph}$ are the normalised energy and differential energy density of the target photons (see equation ~\ref{nph}), $\tilde{\gamma}_\mathrm{-}=\frac{E_\mathrm{-}}{m_\mathrm{e}c^2}$ is the Lorentz factor of the produced electrons in the rest frame of the relativistic proton, $\tilde{p}_{-}=\sqrt{\tilde{\gamma}_\mathrm{-} - 1}$ is the momentum modulus of these electrons normalised by $m_\mathrm{e}c$, and 
$\xi = (\gamma_\mathrm{p}\tilde{\gamma}_{-}- \gamma_\mathrm{e})/(\gamma_\mathrm{p}\tilde{p}_{-})$.

\subsection{
Emission from hadronic cascades}

In the present emission model, we account for electromagnetic cascades initiated by electrons, positrons, and gamma-rays from the photo-pion channel, as well as from 
electron-positron pairs from the Bethe-Heitler process.
Such cascades proceed by the radiative cooling of secondary electrons and positrons whose radiation can be highly energetic to annihilate from the interaction with the soft-photon fields within the emission region.
The photon-photon annihilation creates further electron-positron pairs which cool by radiation and so on.

Given the number density (erg$^{-1}$ cm$^{-3}$) of the high and soft photon fields  
$n^\mathrm{h}$ and $n^\mathrm{soft}$, respectively,
we compute the injection rate $q_\mathrm{ee}^{\gamma\gamma}$  of electron-positron pairs 
(erg$^{-1}$ s$^{-1}$ cm$^{-3}$)
due to photon-photon annihilation following the approach of \cite{Agaronyan_1983}:
\begin{align}
\nonumber
q_\mathrm{ee}^{\gamma\gamma}(E_\mathrm{e})& = 
2\frac{3}{32} \frac{c\sigma_\mathrm{T}}{m_\mathrm{e}c^2}
\int_{\gamma_\mathrm{e}}^{\infty}d\xi_1
\frac{\tilde{n}_1(\xi_1)}{\xi_1^3}
\int_{\frac{\xi_1}{4\gamma_\mathrm{e}(\xi_1 -\gamma_\mathrm{e})}}^{\infty}d\xi_2
\frac{\tilde{n}_2(\xi_2)}{\xi_2^2} \\
\nonumber
&\times\left[
\frac{4\xi_1^2}{\gamma_\mathrm{e}(\xi_1 -\gamma_\mathrm{e})} \ln\left\{\frac{4\xi_2\gamma_\mathrm{e}(\xi_1 -\gamma_\mathrm{e})}{\xi_1}\right\} - 8\xi_1\xi_2 \right. \\
 &\left.+\frac{2(2\xi_1\xi_2-1)\xi_1^2}{\gamma_\mathrm{e}(\xi_1-\gamma_\mathrm{e})}
-\left(1-\frac{1}{\xi_1\xi_2}\right)
\frac{\xi_1^4}{\gamma_\mathrm{e}^2(\xi_1 -\gamma_\mathrm{e})^2}
\right],
\label{qeegg}
\end{align}
where $\gamma_\mathrm{e}=E_\mathrm{e}/(m_\mathrm{e}c^2)$,
$\tilde{n}_2 = m_\mathrm{e} c^2 n^\mathrm{soft}$,  $\tilde{n}_1 = m_\mathrm{e} c^2 n^\mathrm{h}$, and
$\xi_1=\epsilon_1/(m_\mathrm{e}c^2)$ and $\xi_2=\epsilon_2/(m_\mathrm{e}c^2)$ are the 
high and low energy photons, respectively, normalised by the electron rest mass energy. This approach to obtain the pair injection is valid
when $\epsilon_2<<m_\mathrm{e}c^2\lesssim \epsilon_1$, which is the case of the blazar emission model discussed in this paper.

We use eqs. (\ref{nph_gral}-\ref{I_eps}) to compute the high energy photon-field
employing the specific intensity:
\begin{equation}
I^\mathrm{h}_\mathrm{\epsilon} = r_\mathrm{rec} j^\mathrm{h}_\epsilon
\left[1 - \exp\left\{-\tau_{\gamma\gamma}(\epsilon)\right\}\right]
/\tau_{\gamma\gamma(\epsilon)} 
\label{I_h_eps}
\end{equation}
where $j^\mathrm{h}_\epsilon$ is the emission coefficient of the high energy photons and 
$\tau_{\gamma\gamma}$ their optical depth to gamma-ray absorption (given by eq. \ref{tau_gg}).
In all cases, we employ the photon distribution given by eq. \ref{nph} as the low energy photon field density.

To calculate the emission of the injected electron/positrons from photon-photon annihilations, the decay of charged pions, and
the Bethe-Heitler process, 
we consider the electron energy distribution $N_\mathrm{e}$ (erg$^{-1}$ cm$^{-3}$) as the
stationary solution of the particle transport equation \citep{Ginzburg_1964, Romero2010}:  
\begin{equation}
N_\mathrm{e}(E_\mathrm{e}) = 
\frac{
\int^{\infty}_{E_\mathrm{e}} dE'_\mathrm{e} \,q_\mathrm{e,inj}(E'_\mathrm{e})
\exp\left\{-\int_{E_\mathrm{e}}^{E'_\mathrm{e}}
\frac{dE''_\mathrm{e}}{\mid P_\mathrm{e}(E''_\mathrm{e})t_\mathrm{esc}(E''_\mathrm{e})\mid} \right\}
}{\mid P_\mathrm{e}(E_\mathrm{e})\mid}.
\label{Ne_ginzburg}
\end{equation}
In this equation, $q_\mathrm{e,inj}$ is the corresponding electron injection rate (given by eqs. \ref{q_ell}, \ref{q_BH}, or \ref{qeegg}),
$P_\mathrm{e}$ the power of their radiative cooling, and 
$t_\mathrm{esc}$ is the electron escaping time which can be considered as the minimum among the advection  and diffusion time scales.
In this work, we perform our calculations neglecting the
argument of the exponential function in the solution given by eq. \ref{Ne_ginzburg},
which is a valid assumption as long as  $t_\mathrm{adv}$ and $t_\mathrm{diff}$ are much 
longer than the electron cooling time scales. 
We find that secondary electrons/positrons
(which are in general more energetic than primary electrons) have indeed cooling times negligible when compared to the corresponding advection time 
for the radiation solutions derived in this paper. 
The emission spectrum produced by these secondary electron and positrons is then computed as
\begin{equation}
L^\mathrm{syn,e\pm}_{\epsilon}= 
V_\mathrm{b} 4\pi j^\mathrm{syn}_{\epsilon}
\left[1- \exp\{- \tau_{\gamma \gamma}(\epsilon)\} \right]
/\tau_{\gamma \gamma} (\epsilon), 
\label{L_epm}
\end{equation}
where $\tau_{\gamma\gamma}$ is given by eq. (\ref{tau_gg}) and $j^\mathrm{syn}_{\epsilon}$ is given by eq. (\ref{j_eps_syn}).

In our emission model, we account for the contribution to the blazar spectrum of
EM cascades triggered by neutral pion ($\pi^0$) decay, charge pion ($\pi^\pm$) decay, and B-H pair production, as stressed.
The cascade from the $\pi^0$ channel initiates with the injection of gamma-rays for which we employ eq. \ref{I_h_eps} using the emission coefficient $j^{h}_\epsilon = \frac{\epsilon}{4\pi} q_\gamma$ and $q_\gamma$
obtained from eq. \ref{q_ell}.
The cascade from the $\pi^{\pm}$ channel initiates with the injection of electrons and positrons for which we employ eq. \ref{Ne_ginzburg}, using the electron injection rate $q_\mathrm{e, inj} = q_\mathrm{e+}+q_\mathrm{e-}$ (with $q_{e+}$ and $q_{e-}$ given by eq. \ref{q_ell}).
The cascade from the B-H channel initiates with the injection of electron-positron pairs, and in this case we apply
eq. (\ref{Ne_ginzburg}), with $q_\mathrm{e, inj} = q^\mathrm{B-H}_\mathrm{e\pm}$ from eq. \ref{q_BH}).

\subsection{Observed fluxes}
Our emission model assumes the trajectory of the emitting blob as quasi-aligned to the line of sight with an angle 
$\theta_\mathrm{b}<1/\Gamma_\mathrm{j}$.
Under this condition, we calculate the observed flux (erg s$^{-1}$ cm$^{-2}$) of an emission component
at the observed energy $E$, as
\begin{equation}
E F_{E} = \frac{\Gamma^4_\mathrm{j} \epsilon L_{\epsilon}}{4\pi D^2_\mathrm{L}} 
\exp\left\{-\tau^\mathrm{eg}_{\gamma\gamma}(E,z)\right\},
\label{obs_flux}
\end{equation}
where $\epsilon$ and $E = \epsilon \Gamma_\mathrm{j} /(1+z)$ are the emission energies at the source and the observer frames, respectively, 
$z$ and $D_\mathrm{L}$ are the redshift and luminosity distance of the source, respectively,
and $L_\epsilon$ is the differential luminosity of the emission component in the source frame (see eqs.
\ref{Lsyn}, \ref{L_e_C}, \ref{Lnus}, and \ref{L_epm}).
We assume the source with a redshift and luminosity distance of
$z=0.34$ and $D_\mathrm{L}= 1750$ Mpc,
respectively \citep{neutrinosmm}.
The attenuation of the blazar emission due to the extragalactic light (EBL) is accounted  by the attenuation coefficient in the exponential factor in the RHS of eq. \ref{obs_flux},
which is relevant for the emission components producing photons at GeV-TeV energies. We calculate this coeffiecient using 
the gammapy package \citep{Donath_2023},
adopting the EBL absorption model of \cite{Dominguez_2011}





\bsp	
\label{lastpage}
\end{document}